\documentclass{aa}  

\usepackage{graphicx}
\usepackage{txfonts}
\usepackage{amsmath}
\usepackage{bm}
\usepackage{hyperref}
\usepackage{xcolor}
\usepackage{placeins}
\newcommand{\micron}{$\rm \mu$m}

\begin{document}

   \title{The disk of FU Orionis viewed with MATISSE/VLTI\thanks{Based on observations collected at the European Southern Observatory under ESO programs 0104.C-0782(B), 0104.C-0016(D), 0106.C-0501(D), and 0106.C-0501(F).} }

   \subtitle{First interferometric observations in $L$ and $M$ bands.}

   \author{F.~Lykou \inst{1}
          \and
          P.~\'Abrah\'am\inst{1,2}
          \and
          L.~Chen\inst{1}
          \and
          J.~Varga\inst{3,1}
          \and
          \'A.~K\'osp\'al\inst{1,2,4}
          \and
          A.~Matter\inst{5}
          \and
          M.~Siwak\inst{1}
          \and
          Zs.~M.~Szab\'o\inst{6,7,1}
          \and
          Z.~Zhu\inst{8}
          \and
          H.~B.~Liu\inst{9}
          \and
B.~Lopez \inst{5}
\and
F.~Allouche \inst{5}
\and
J.-C. Augereau\inst{10}
\and
P.~Berio \inst{5}
\and
P.~Cruzal\`ebes \inst{5}
\and
C.~Dominik \inst{11}
\and
Th.~Henning \inst{4}
\and
K.-H.~Hofmann \inst{6}
\and
M.~Hogerheijde \inst{3,11}
\and
W.~J.~Jaffe \inst{3}
\and
E.~Kokoulina \inst{5}
\and
S.~Lagarde \inst{5}
\and
A.~Meilland \inst{5}
\and
F.~Millour \inst{5}
\and
E.~Pantin \inst{12}
\and
R.~Petrov \inst{5}
\and
S.~Robbe-Dubois \inst{5}
\and
D.~Schertl \inst{6}
\and
M.~Scheuck \inst{4}
\and
R.~van Boekel \inst{4}
\and
L.~B.~F.~M.~Waters \inst{13,14}
\and
G.~Weigelt \inst{6}
\and
S.~Wolf \inst{15}
          }

   \institute{Konkoly Observatory, Research Centre for Astronomy and Earth Sciences, E\"otv\"os Lor\'and Research Network (ELKH), Konkoly-Thege Mikl\'os \'ut 15-17, 1121 Budapest, Hungary
     \email{foteini.lykou@csfk.org}
         \and
         ELTE E\"otv\"os Lor\'and University, Institute of Physics, P\'azm\'any P\'eter s\'et\'any 1/A, 1117 Budapest, Hungary
         \and
         Leiden Observatory, Leiden University, P.O. Box 9513, 2300 RA, Leiden, The Netherlands
         \and
         Max-Planck-Institut f\"ur Astronomie, K\"onigstuhl 17, D-69117 Heidelberg, Germany 
         \and
         Laboratoire Lagrange, Universit\'e C\^ote d’Azur, Observatoire de la C\^ote d’Azur, CNRS, Boulevard de l’Observatoire, CS 34229, 06304, Nice Cedex 4, France
                \and
                Max-Planck-Institut f\"ur Radioastronomie, Auf dem H\"ugel 69, 53121 Bonn, Germany
                \and
                Scottish Universities Physics Alliance (SUPA), School of Physics and Astronomy, University of St Andrews, North Haugh, St Andrews, KY16 9SS, UK
        \and
         Department of Physics and Astronomy, University of Nevada, Las Vegas, 4505 S. Maryland Parkway, Las Vegas, NV 89154, USA
         \and
         11F of AS/NTU Astronomy-Mathematics Building, No.1, Sect. 4, Roosevelt Rd, Taipei 10617, Taiwan, R.O.C.
         \and
         Univ. Grenoble Alpes, CNRS, IPAG, 38000 Grenoble, France
         \and
         Anton Pannekoek Institute for Astronomy, University of Amsterdam, Science-Park 904, 1098 XH, Amsterdam, The Netherlands
         \and
         AIM, CEA, CNRS, Universit\'e Paris-Saclay, Universit\'e Paris Diderot, Sorbonne Paris Cit\'e, 91191 Gif-sur-Yvette, France
         \and
         Institute for Mathematics, Astrophysics and Particle Physics, Radboud University, PO Box 9010, MC 62, 6500 GL, Nijmegen, The Netherlands
         \and
        SRON Netherlands Institute for Space Research, Niels Bohrweg 4, 2333 CA, Leiden, The Netherlands 
        \and
        Institut f\"ur Theoretische Physik und Astrophysik, Christian-Albrechts-Universit\"at zu Kiel, Leibnizstraße 15, 24118, Kiel, Germany 
                }

   \date{Received Dec 01, 2021; accepted Apr 20, 2022}

 
  \abstract
   {} 
   {We studied the accretion disk of the archetypal eruptive young star FU Orionis with the use of mid-infrared interferometry, which enabled us to resolve the innermost regions of the disk down to a spatial resolution of 3~milliarcseconds (mas) in the $L$ band, that is, within 1~au of the protostar.}
   {We used the interferometric instrument MATISSE/VLTI to obtain  observations of FU Ori's disk in the $L$, $M$, and $N$ bands with multiple baseline configurations. We also obtained contemporaneous photometry in the optical ($UBVRIr'i'$; SAAO and Konkoly Observatory) and near-infrared ($JHK_s$; NOT). Our results were compared with radiative transfer simulations modeled  by {\sc radmc-3d}.}
   {The disk of FU Orionis is marginally resolved with MATISSE, suggesting that the region emitting in the thermal infrared is rather compact. 
   An upper limit of $\sim1.3\pm0.1$~mas (in $L$) can be given for the diameter of the disk region probed in the $L$ band, corresponding to 0.5 au at the adopted {\it Gaia} EDR3 distance. This represents the hot, gaseous region of the accretion disk. The $N$-band data indicate that the dusty passive disk is silicate-rich. Only the innermost region of said dusty disk is found to emit strongly in the $N$ band, and it is resolved at an angular size of $\sim5$~mas, which translates to a diameter of about 2 au. The observations therefore place stringent constraints for the outer radius of the inner accretion disk.  
   Dust radiative transfer simulations with {\sc radmc-3d} provide adequate fits to the spectral energy distribution from the optical to the submillimeter and to the interferometric observables when opting for an accretion rate $\dot{M}\sim 2\times 10^{-5}$~$M_\sun$~yr$^{-1}$ and assuming $M_*=0.6$~$M_\sun$. Most importantly, the hot inner accretion disk's outer radius can be fixed at 0.3~au. The outer radius of the dusty disk is placed at 100~au, based on constraints from scattered-light images in the literature. The dust mass contained in the disk is $2.4\times10^{-4}$~$M_\sun$, and for a typical gas-to-dust ratio of 100, the total mass in the disk is approximately 0.02~$M_\sun$. We did not find any evidence for a nearby companion in the current interferometric data, and we tentatively explored the case of disk misalignment. For the latter, our modeling results suggest that the disk orientation is similar to that found in previous imaging studies by ALMA. Should there be an asymmetry in the very compact, inner accretion disk, this might be resolved at even smaller spatial scales ($\leq1$~mas).
   }
   {}

   \keywords{Stars: individual: FU Ori -- protoplanetary disks -- circumstellar matter -- infrared: stars -- techniques: interferometric -- radiative transfer
               }

   \maketitle
%

\section{Introduction}
FU Orionis is the archetype of the FUor class of young stellar objects (YSOs), which experience eruptive events initiated by increased accretion (order of $\sim10^{-4}$~$M_\sun$ yr$^{-1}$) of material from their circumstellar disks onto the stellar surface \citep{1977ApJ...217..693H,1996ARA&A..34..207H}. These eruptions subsequently heat the material near the protostar (i.e., the inner accretion disk) and result in a brief rise in brightness with a subsequent decay. For the case of FU Orionis, the eruption increased its brightness by $\Delta B \sim 6$ magnitudes within a single year \citep[1936-37;][]{1939BHarO.911...41H,1966VA......8..109H,1988ApJ...325..231K}. According to \citet{2000ApJ...531.1028K}, the $B$-band brightness has been declining slowly ever since by approximately 0.015 mag per year. Not more than 20 other stars have been found to belong to the FUor class thus far \citep[e.g.,][]{connelley2018}. It is unclear whether FUors are an entirely unique class of YSOs or a common step in the evolutionary path for all YSOs \citep[e.g.,][]{2009apsf.book.....H}.

FU~Ori is located at a distance of $402.3^{+3.0}_{-3.7}$\,pc according to {\it Gaia} Early Data Release 3 \citep[EDR3;][]{2021AJ....161..147B}.
It is thought to be a young low-mass  object based on theoretical estimates \citep[e.g., 0.6~$M_\sun$;][]{2020ApJ...889...59P, 2020MNRAS.495.3494Z}. The star is surrounded by a circumstellar disk \citep{1985ApJ...299..462H} and, on a larger scale, by a reflection nebulosity \citep{1954ZA.....35...74W, 1966VA......8..109H}. The properties of the circumstellar accretion disk around FU~Orionis, such as its orientation, accretion rate, and composition, have been extensively studied in the past. The majority of earlier works \citep[e.g.,][]{calvet1991,1996ARA&A..34..207H, henning1998, 2000ApJ...531.1028K, 2007ApJ...669..483Z} provided parametric studies of the disk based on fitting the spectral energy distribution (SED) and spectroscopic observations with accretion disk models; this continued until the arrival of high-spatial-resolution observations, which allowed the disk to be directly detected. 

Scattered-light images in the near-infrared reveal an arc-shaped stream of material appearing at about 0.3\arcsec\ directly east of FU~Ori, with additional gaps in the northern and western directions and a tentative suggestion of a jet ejection \citep{2016SciA....2E0875L,2018ApJ...864...20T, 2020ApJ...888....7L}. Furthermore, \citet{2020ApJ...889...59P} suggested that the gaseous component of FU~Ori's disk is in fact in Keplerian rotation and possibly related to the arc-shaped stream seen in scattered light.

FU~Orionis is a member of a wide binary system. The companion star, FU Ori S, was located approximately $\sim$0.5\arcsec\ away at a position angle (PA) of $\sim161\degr$ in 2002 \citep{2004ApJ...601L..83W}. Its presence has been confirmed by many follow-up imaging studies \citep[e.g.,][]{2004ApJ...608L..65R}, and it is believed to be a pre-main-sequence K star. Far-millimeter continuum emission suggests that the companion also hosts its own circumstellar disk, and that both disks are aligned in similar directions \citep{2015ApJ...812..134H,2017A&A...602A..19L,2019ApJ...884...97L, 2020ApJ...889...59P}.

The introduction of infrared interferometry since the late 1990s has allowed an even closer inspection of the disk, as multiwavelength studies have found the apparent size of the accretion disk's innermost hot region to be less than 2~au \citep{1998ApJ...507L.149M, 2005A&A...437..627M, 2006ApJ...648..472Q, 2009ApJ...700..491M, 2019ApJ...884...97L, 2021A&A...646A.102L}. These earlier studies explored the case of a third companion near FU~Orionis, although a concrete detection is yet to be made. More recently, \cite{2019ApJ...884...97L} and \cite{2021A&A...646A.102L} measured near-zero closure phases, which indicate an inclined, centro-symmetric distribution and would exclude the presence of a nearby companion. Nevertheless, \cite{2021A&A...646A.102L} proposed an upper limit of about 1.3\% for the $H$-band flux ratio for a subsolar companion with a separation between 0.5 and 50 milliarcseconds (mas), or between 0.5 and 20 au for our adopted {\it Gaia} distance.

Nearly all interferometric studies find different geometric properties, that is, inclination and PA, for the FU Orionis disk. Since these measurements were inferred at different wavelengths, from the near-infrared to the radio regimes, it is worth questioning whether individual components of the circumstellar disk may be misaligned. The most precise measurements thus far were made via direct imaging of the disk in submillimeter wavelengths \citep{2020ApJ...889...59P}. Since the geometric properties are a crucial input in models, the results of disk simulations (such as inferring the stellar radius and mass, the disk's inner and outer radius, and the mass accretion rate) can often vary. For the mass accretion rate in particular, such variations are found to be on the order of $\sim 10^{-4}$ $M_\sun$ yr$^{-1}$. For instance, the boundary between the hot inner disk and the passive, cooler component is found at $0.76\pm0.35$ au with the outer disk radius fixed at 7.7 au by \cite{2021A&A...646A.102L}, assuming a distance of 416 pc; this may be in agreement with \cite{2019ApJ...884...97L}, suggesting that submillimeter and centimeter emission originates within 10 au radii.
On the other hand, \cite{2007ApJ...669..483Z} place the outer radius of the hot, inner disk at $\sim0.5$ au (at a distance of 500 pc), with a not well-confined truncation radius of $\sim1$ au between the inner and the dusty disk, as it is not clear which disk component dominates in emission in the 4 - 8~\micron\ region \citep[see also ][]{2008ApJ...684.1281Z}. Recently, \citet{2021ApJ...923..270L} found that the disk's mid-plane is dominated by millimeter-sized grains and posited that a temporal change in the source's brightness may be the result of dynamical changes in the disk.

In this work we present the first observations of FU~Orionis with the Multi AperTure mid-Infrared SpectroScopic Experiment (MATISSE) instrument of the Very Large Telescope Interferometer (VLTI) obtained in three bands, namely $L$, $M$ (3 - 5 $\rm \mu m$), and $N$ (8 - 13 $\rm \mu m$) \citep{matisse, matisse2, matisse1, matisse4, lopez2021matisse}. They were complemented by nearly simultaneous photometric observations in the optical and near-infrared from the South African Astronomical Observatory (SAAO), the Nordic Optical Telescope (NOT), and the Konkoly Observatory. Our findings are presented in Sects.~\ref{sec:obs} and \ref{sec:results}. The interferometric results are compared with analytical and radiative transfer models for the inner and outer parts of the accretion disk, respectively, and can be found in Sect.~\ref{sec:simulations}, which is followed by a discussion on our findings in Sect.~\ref{sec:discuss}. Our conclusions are described in the final section.

\section{Observations}\label{sec:obs}

\subsection{MATISSE/VLTI long-baseline interferometry}\label{sec:matisse}
Guaranteed Time Observations (GTO) were obtained at the European Southern Observatory with the 4-beam recombiner instrument MATISSE \citep{matisse,matisse2,matisse1,matisse4,lopez2021matisse} on the VLTI over five epochs (Table~\ref{tab:log}) with different baseline configurations with the 8.2 m unit telescopes (UTs) and the 1.8 m auxiliary telescopes (ATs). These configurations produce baseline coverage between 40 and 130 meters, with angular resolutions of $\sim$ 3 - 9 mas in $L$ and $\sim$ 8 - 26 mas in $N$, respectively. Hereafter, the term ``baseline'' will always refer to the ``projected'' baselines (length and/or PA) on the sky. 

Here, we present results from two epochs (i.e., epochs 4 and 5, Table~\ref{tab:log}) for which we obtained simultaneous photometric observations (see the following sections). A brief description of the data quality of all epochs is given in Appendix~\ref{sec:log}.

We opted for low-spectral resolution ($R\sim30$) in $L$ and $M$ and for high-spectral resolution ($R\sim220$) in the $N$ band in epoch 4 with the UTs. These observations were obtained in the standard hybrid mode, which utilizes the simultaneous photometry (SiPhot) mode in the $L$ band and the high sensitivity (HighSens) mode in the $N$ band. In the former, photometry is recorded simultaneously to the interferometric fringes in a separate channel after splitting the incoming beam, while in the latter mode the photometry is collected after measuring the interferometric fringes. A similar spectral setup was chosen for epoch 5 (ATs), that is, low-spectral resolution in $LM$ and high in $N$. However, these data sets were obtained in the new GRAVITY for MATISSE (GRA4MAT) mode \citep{lopez2021matisse}, which is designated for faint targets. In this mode, the GRAVITY/VLTI instrument is used as a fringe-tracker, thereby providing improved sensitivities for the ATs and increasing the spectral coverage. This was essential for FU Ori since we found that, for such a faint target, observations in the standard hybrid mode and the ATs were very difficult (cf. Appendix~\ref{sec:log}).

The observations benefited from excellent atmospheric conditions with an average seeing as low as 0.5\arcsec\ (cf. Table~\ref{tab:log}). The atmospheric transfer function was estimated with Mid-infrared stellar Diameters and Fluxes compilation Catalogue \citep[MDFC; ][]{2019MNRAS.490.3158C} calibrators HD48433 (K0III), HD28413 (K4III), HD47886 (M1III), HD37160 (G9.5III), and HD31767 (K2II). The data sets were reduced with the standard MATISSE data reduction software (DRS) pipeline (versions 1.5.2 and 1.6) and designated MATISSE {\sc python} tools\footnote{\url{https://gitlab.oca.eu/MATISSE/tools/-/wikis/home}}. Since FU Ori is quite faint in all bands, that is, below the minimum suggested from MATISSE specifications\footnote{\url{https://www.eso.org/sci/facilities/paranal/instruments/matisse/inst.html}}, when reducing the $N$-band UT data we opted to use a larger spectral bin (16 pixels) as opposed to the standard method (7 pixels). This averaging minimizes the noise in the correlated fluxes and in the closure phases, in order to reveal any deviations from spherical symmetry. A detailed description of the data calibration for MATISSE has been presented in previous works \citep[e.g.,][]{2021A&A...647A..56V}; therefore, we refrain from repeating it here. The interferometric products are presented in Sect.~\ref{sec:results}.

\subsection{Optical and infrared photometry}\label{sec:phot}

With the goal to construct an SED close in time to MATISSE interferometric observations, we collected optical and near-infrared photometry in the period from January 2021 to February 2021. A description of these observations follows below.

\subsubsection{Nordic Optical Telescope}\label{sec:not}

Photometric observations were obtained with the NOT near-infrared camera and spectrograph \citep[NOTCam;] []{2000SPIE.4008..714A} on 25 January and 18 February 2021 (program 62-410, PI: F.~Lykou), with the high-resolution (HR) imaging mode in standard $J$, $H$, and $Ks$ filters. The detector array ($1024\times1024$ pixels) offers a field-of-view of 80\arcsec$\times$80\arcsec\ in HR mode with a plate scale of 0.078\arcsec/pixel. To avoid saturation due to the brightness of FU~Ori in the near-infrared, a 5~mm pupil mask was introduced that reduced the transmission down to 10\%. The average seeing on both nights was $\leq1.2$\arcsec.

FU~Orionis exposures were bracketed by two comparison stars (2MASS J05452885+0901452, 2MASS J05451389+0904443) to assist in photometric calibration. In the first epoch, all three stars were exposed for 4.5 sec per dither in a five-point dithering pattern (read-read-reset mode) in all three filters. Due to final low signal-to-noise of the faint calibrator on the first epoch (January), we repeated the observations in February using the same camera setup but with a nine-point dithering pattern and 3.6 sec per dither with the ramp-sampling readout mode.

A median sky frame was subtracted from each science frame, and a differential flat-field was used to correct for pixel-to-pixel variations. Photometry was computed in the Two Micron All Sky Survey (2MASS) system for each dithering position against both comparison stars in varying aperture radii between 10 and 50 pixels, and it stabilized at 32 pixels ($\sim2.5$\arcsec). FU~Ori's brightness was found to be consistent between the two epochs at 6.90, 6.21, and 5.65 mag in $J$, $H$, and $Ks$, respectively, with a typical uncertainty $\delta m\sim0.03$~mag. The NOTCam photometry is included in Table~\ref{tab:sed}.


\subsubsection{South African Astronomical Observatory}\label{sec:saao}
Optical photometric measurements were obtained at SAAO simultaneously with the NOT observations (25 January 2021; P.I.: M.~Siwak, program ID: Siwak-2020-05-40-inch-317). We utilized the Sutherland Highspeed Optical Cameras \citep[SHOC;][]{2013PASP..125..976C} with Bessel $UBVRI$ filters, mounted on the 1-meter Lesedi telescope. The instrument offers a 5.72\arcmin$\times$5.72\arcmin\ field-of-view with a 0.335\arcsec/pixel plate scale; however, 2x2 binning was used to match to the seeing conditions. The secondary standards GSC~00714-00203 and GSC~00715-00188 were used as calibrators \citep{2018A&A...618A..79S}. The photometry is tabulated in Table~\ref{tab:sed}, where $UBV$ are in the Bessel system, and $R_cI_c$ in the Cousins system.


\subsubsection{Konkoly Observatory}\label{sec:rc80}
Photometry was obtained with the fully automated Astro Systeme Austria (ASA) AZ800 alt-azimuth, direct-drive, 80-centimeter Ritchey–Chr\'tien (RC) telescope at the Piszk\'estető Observatory, Hungary, on 13--18 February 2021. We used a 2048$\times$2048 pixel FLI~PL230~CCD camera with the E2V~CCD230-42 detector. The optical setup with an effective focal distance of $f=5600\,{\rm mm}$ yields a pixel scale of 0.55\arcsec\ and a field-of-view of 18.8\arcsec$\times$18.8\arcsec. We obtained three images per night with Bessel $B$, $V$ and Sloan $r'$ and $i'$ filters. 

A standard data reduction method was followed by applying bias, dark, and flat-field corrections. Aperture photometry was obtained for both the science target (FU Ori) and about 30 nearby, comparison stars using an aperture radius of 10 pixels (5.5\arcsec) and a sky annulus of 10-15 pixels (5.5\arcsec\ - 8.25\arcsec). Photometric calibration was done by fitting a color term using the magnitudes and colors of the comparison stars from the AAVSO Photometric All-Sky Survey (APASS) DR9 catalog \citep{henden2016}. The resulting photometry is presented in Table~\ref{tab:sed}, where $BV$ are in the Bessel system and $r'i'$ in the Sloan system. The typical uncertainty of the photometry is 0.01\,mag.

\begin{figure}[htbp]
\centering
        \includegraphics[width=\columnwidth]{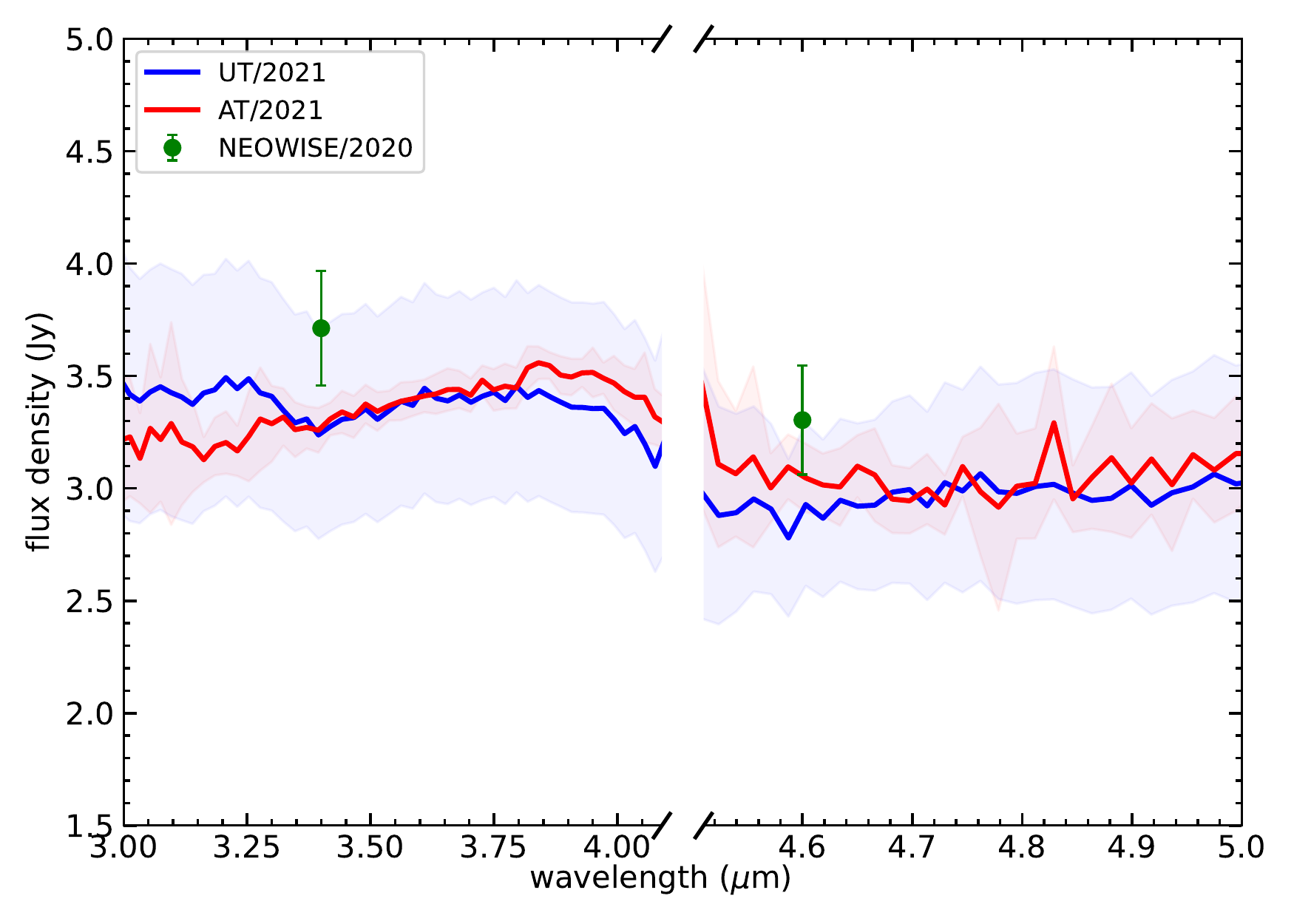}\\
    \includegraphics[width=\columnwidth]{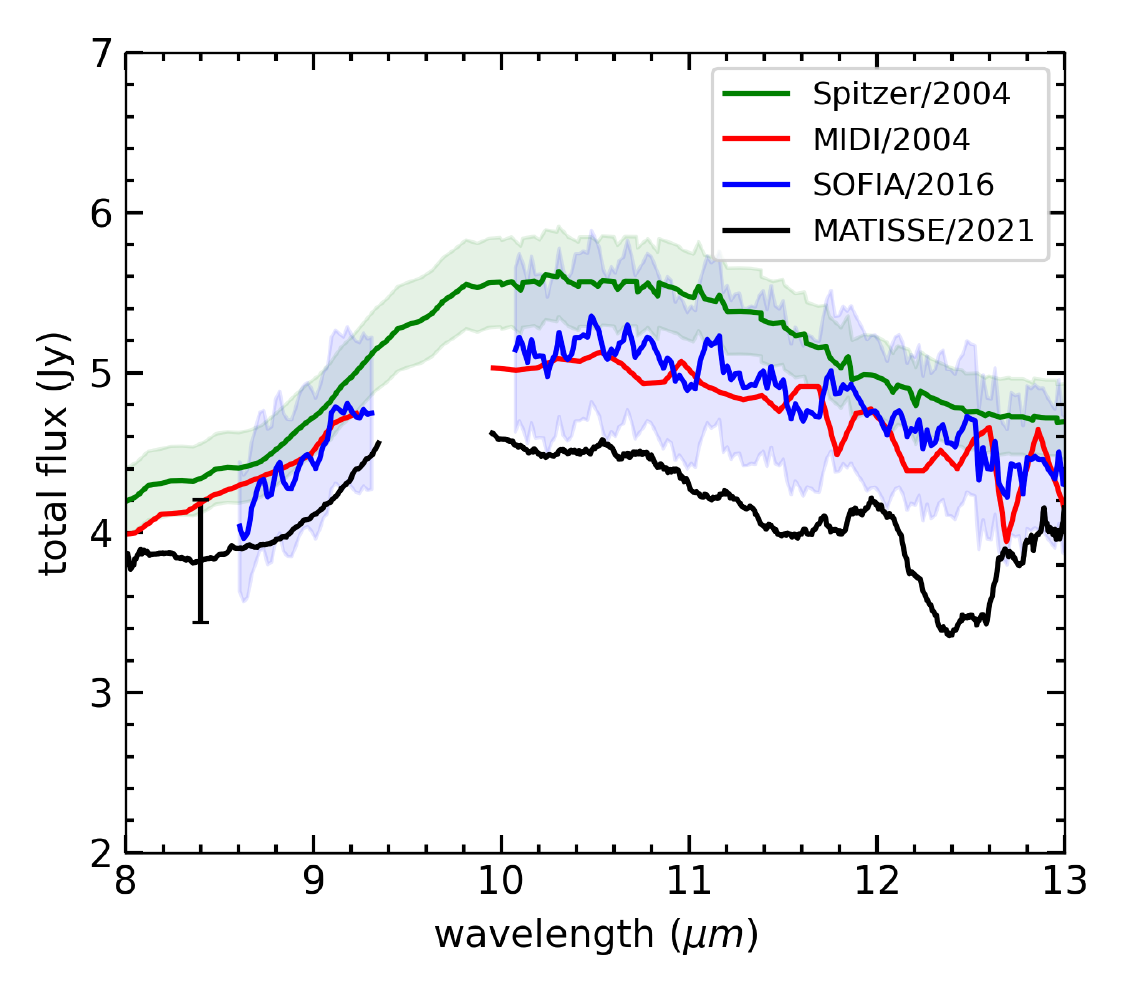}
        \caption{MATISSE total flux spectra. {\bf Top:} $L$ and $M$ band from the UTs (blue) and the ATs (red; GRA4MAT) from epochs 4 and 5, respectively (Table~\ref{tab:log}). NEOWISE photometry from 2020 (corrected for saturation) is shown for comparison (green). {\bf Bottom: } $N$-band spectrum (black; UTs) against earlier spectra from {\it Spitzer} \citep[green; ][]{2006ApJ...648.1099G}, MIDI \citep[red; ][]{2006ApJ...648..472Q}, and SOFIA \citep[blue; ][]{green2016}. The flux uncertainty of MATISSE is about 10\% for the case of FU~Ori (error bar in the lower panel). Spectral regions where atmospheric transmittance hinders terrestrial observations are cut off from the spectra.
        }
        \label{fig:LMNflux}
\end{figure}


\subsection{Interstellar extinction}\label{sec:extinction}
The neutral hydrogen column density, $N_H$, in a region within 0.1\degr\ of FU~Ori's location is $1.99\times10^{21}$ atoms/cm$^2$ based on HI 4-PI Survey (HI4PI) data \citep{hi4pi}. Adopting the calculation of \citet{2009MNRAS.400.2050G} where $N_H = (2.21\pm0.09)\times10^{21}\,A_V$, we find a low interstellar extinction of $A_V=0.90\pm0.04$ mag. This should account for foreground extinction within a region of 0.1\degr; however, FU~Ori is located inside the dust cloud B35 and therefore the extinction is suspected to be higher.

Conflicting values have been found in the literature with regard to the measurement of the interstellar extinction. Previous estimates were based on fitting FU~Ori's broadband photometry and/or spectra, and placed $A_V$ between 1.5 \citep{2007ApJ...669..483Z} and 3.2 mag \citep{1966VA......8..109H}. Here, we adopt the value derived from independent methods, and in particular from the 3D Dust Maps of \cite{2019ApJ...887...93G} for the adopted {\it Gaia} distance of $402.3^{+3.0}_{-3.7}$\,pc \citep{2021AJ....161..147B}, that is, $A_V = 1.7\pm0.1$ mag. 


\section{Interferometric data}\label{sec:results}

\subsection{Total flux}
We draw caution on the absolute flux calibration of MATISSE in all three bands. On the one hand, the performance is inherent to the source's brightness itself (cf. Sect.~\ref{sec:matisse}), on the other hand it is dependent on atmospheric conditions. The latter can be assessed through the variability of the transfer function. It should be also optimized when using chopping in the $L$ band to remove the thermal background, as this is affected more by the limited terrestrial atmospheric transmission between 3 and 5~\micron.

Figure~\ref{fig:LMNflux} (top panel) shows the total flux spectra in $LM$ from the UTs (blue) and from the ATs (red). A description of the data quality is given in Appendix~\ref{sec:log}. Overall, the flux levels agree quite well despite being obtained in different modes and baseline configurations, while they are directly comparable to Near-Earth Object Wide-field Infrared Survey Explorer (NEOWISE) photometry obtained almost a year earlier (corrected for saturation; green points). 

The $N$-band total flux spectrum is shown in the bottom panel of Fig.~\ref{fig:LMNflux} (black line). We find that the flux uncertainty of MATISSE with the UTs for epoch 4 is about 10\% for this faint target. This is mostly due to calibration errors and it can become worse beyond 11\micron. Despite the higher sensitivity offered with the UTs on MATISSE, the S/N decreases beyond 11~\micron\ for FU~Ori (cf. Appendix~\ref{sec:log}); therefore, all interferometric products will be noisier at the tail end of the $N$ band. No photometry was obtained in epoch 5.

When compared with earlier spectra from satellite, airborne, and terrestrial telescopes -- {\it Spitzer}, \citep[green line;][]{2006ApJ...648.1099G},  the Stratospheric Observatory for Infrared Astronomy \citep[SOFIA; blue line;  ][]{green2016}, and the  MID-infrared Interferometric (MIDI) instrument \citep[red line; ][]{2006ApJ...648..472Q}, respectively -- we note that the total flux appears to have dimmed by at least 0.5~Jy since 2004. However, due to the flux calibration uncertainties from all instruments, this is not clear. Further discussion on this dimming can be found in Sect.~\ref{sec:sed}.


\subsection{Visibilities and correlated fluxes}\label{sec:V2}

\subsubsection{L and M bands}
\begin{figure*}[htbp]
\centering
                \includegraphics[scale=0.55]{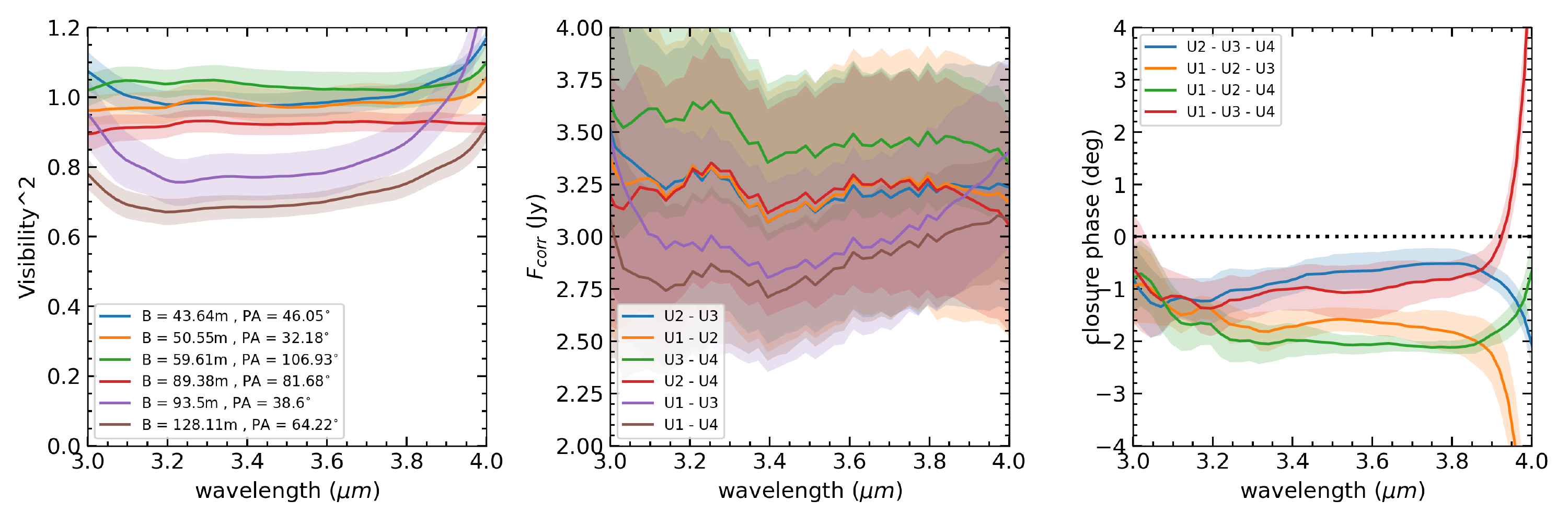}\\
                \includegraphics[scale=0.55]{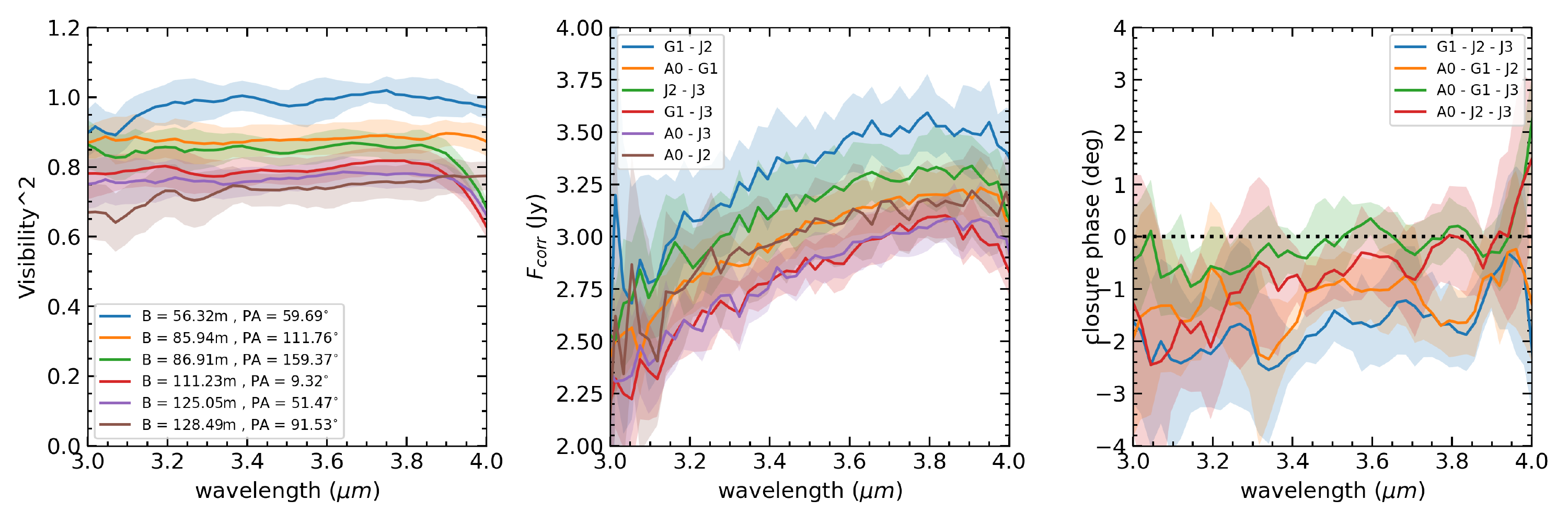}
    \caption{MATISSE squared visibilities, correlated fluxes, and closure phases in the $L$ band. The first row shows data from epoch 4 (UTs) and the bottom row data from epoch 5 (ATs). The visibilities and correlated fluxes are color-coded with respect to the baselines, as shown in the legends, while closure phases have a different color scheme (per baseline triangle). The apparent flaring in the data below 3.2\micron\ and beyond 3.9\micron\ is a known instrumental artifact at the edges of the atmospheric transmission windows.}
    \label{fig:Lplots}
\end{figure*}

\begin{figure*}
    \centering
                \includegraphics[scale=0.55]{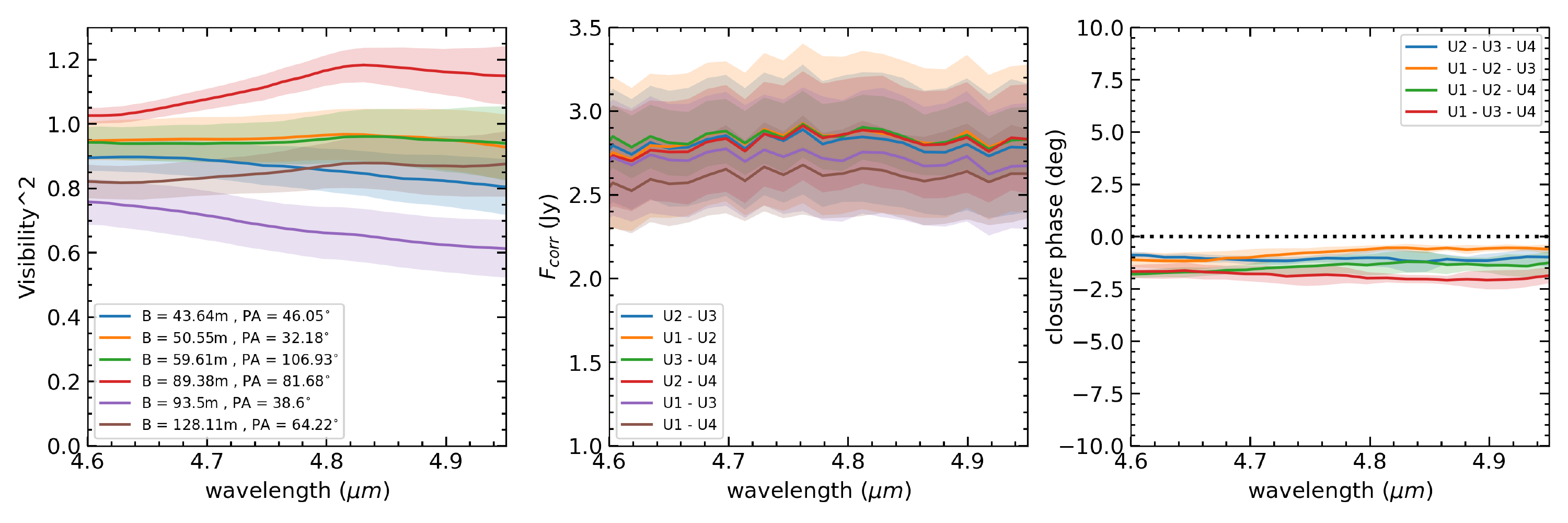}\\
                \includegraphics[scale=0.55]{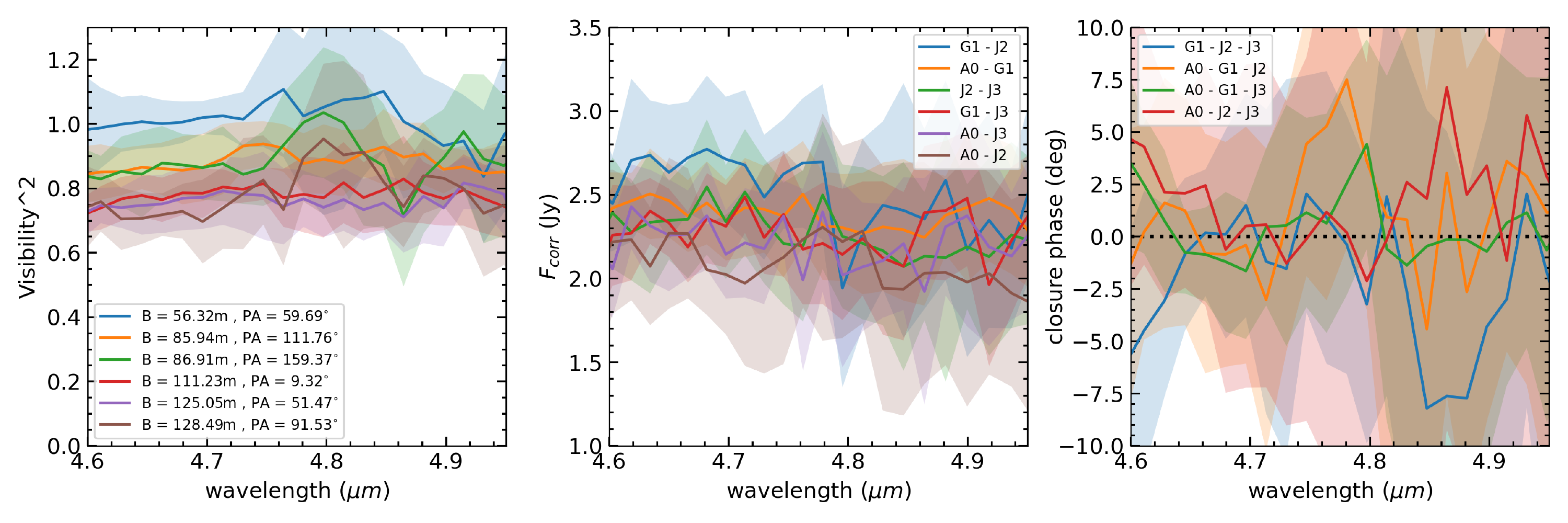}
    \caption{As in Fig.~\ref{fig:Lplots} but for the $M$ band.}
    \label{fig:Mplots}
\end{figure*}

The squared visibilities, $V^2$, and correlated fluxes, $F_{\rm corr}$, in $L$ and $M$ bands are shown in the left and middle columns in Figs.~\ref{fig:Lplots} and~\ref{fig:Mplots}. The top panels are the data from the UTs, and the bottom panels are the data from the ATs. The influence of the terrestrial atmosphere is evident in the abrupt changes in the $L$-band $V^2$ and $F_{\rm corr}$ below 3.2~\micron\ and beyond 3.9~\micron, that is, at the edges of this atmospheric transmission window (see also Appendix~\ref{sec:log} for a further description on data quality). Despite the different observational setups between the two epochs, we find that the results are quite similar in both bands.

In both the $L$ and $M$ bands, the squared visibilities approach unity at the shorter baselines (i.e., $B\leq90$ m or else spatial scales $>4$~mas) and thus indicate that FU Ori's inner disk region is almost unresolved at those spatial scales. However, that region is marginally resolved at the longest baselines (spatial scales $<4$~mas) where $0.65 \leq V^2 \leq 0.80$. This suggests that the emitting region in $L$ and $M$ (i.e., the accretion disk) is rather compact and confined within an area with a global size $\leq1.6$~au; therefore, its radius should be smaller than 0.8~au. The correlated fluxes show a similar behavior to the squared visibilities with minute variations between baselines. Overall, the correlated spectra are flat and do not show any of the typical spectral features of FUors (e.g., Br$\alpha$, CO), but this should be expected in this case where low-spectral-resolution setups were used.

\subsubsection{N band}
\begin{figure*}[htbp]
        \centering
                \includegraphics[scale=0.55]{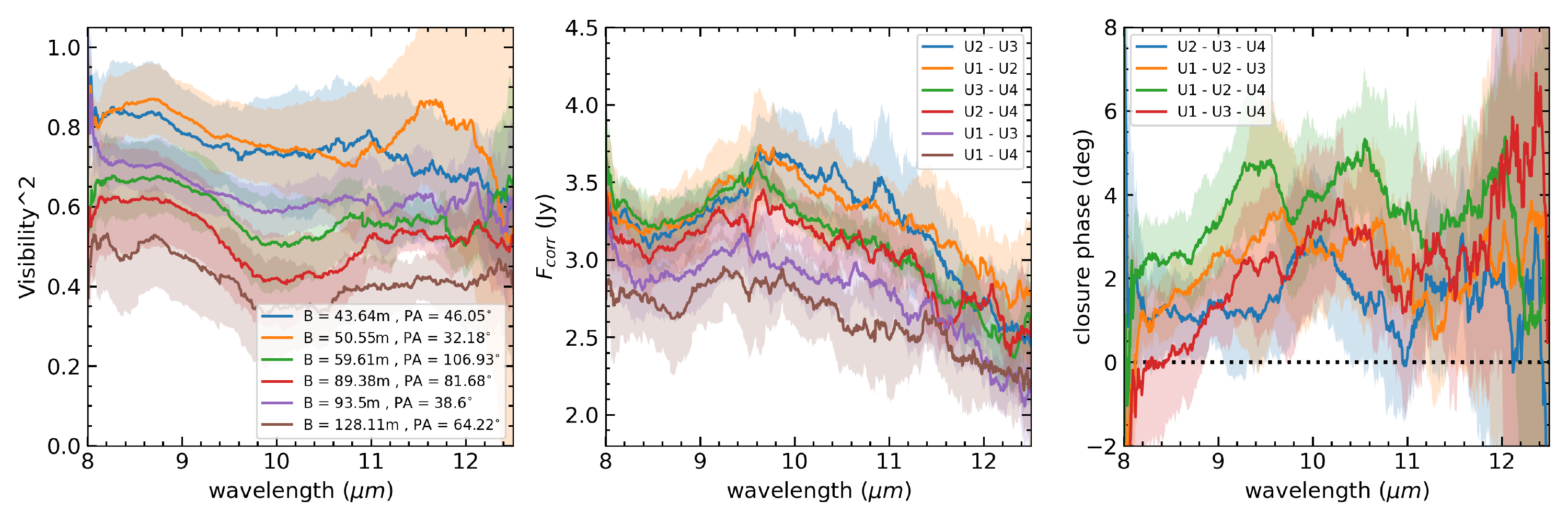}\\
    \caption{As in Fig.~\ref{fig:Lplots} but for $N$-band data of epoch 4. }
    \label{fig:Nplots}
\end{figure*}

The $N$-band data from epoch 4 are shown in Fig.~\ref{fig:Nplots}. FU~Ori is known to have the 10~\micron\ silicate feature in emission \citep[e.g.,][]{2006ApJ...648.1099G,2006ApJ...648..472Q} and that is indeed seen in the MATISSE correlated spectra and the visibilities. Although the sensitivity of the instrument was increased with the introduction of GRA4MAT in epoch 5, the $N$-band brightness of FU Orionis is still below the GRA4MAT sensitivity limit; therefore, the observations suffered from very low S/N and are not presented here. 

The visibilities also indicate that a part of the dusty disk is resolved by MATISSE more at the longest baselines (i.e., $90\leq B \leq 120$~m range). This suggests that the innermost region of the dusty disk is located inside an area of 9~mas in size (at 10~\micron), otherwise within a radius of 2~au from the protostar. However, that area accounts for less than 50\% of the 10~\micron\ emission since the average visibility is $\leq0.5$. The majority of the 10~\micron\ silicate dust emission arises from a larger region of the disk, which can be probed by MATISSE with the shortest UT baselines ($B\sim40$~m). Its size ought to be smaller than roughly 25~mas at 10~\micron, corresponding to an area smaller than 10~au. The physical size of the disk is expected to be much larger, perhaps a few hundred astronomical units, as hinted by near-infrared scattered-light images. However, the extended disk is not seen here by MATISSE because either (a) it could only be probed at shorter baseline configurations (e.g., $B \leq 20$~m), or (b) the extended disk is not a strong emitter in the mid-infrared. Overall, the $N$-band visibilities measured by MATISSE are consistent with those measured by MIDI \citep{2006ApJ...648..472Q}.


\subsection{Closure phases}\label{sec:CPs}

The MATISSE closure phases in $L$, $M$, and $N$ are shown in the rightmost columns in Figs.~\ref{fig:Lplots}, \ref{fig:Mplots}, and \ref{fig:Nplots}. In the $L$ and $M$ bands, the closure phases are consistent between epochs and observational setups, and remain constant within each band with an average value of 1.5\degr. One exception are the $M$-band data of epoch 5 (ATs; GRA4MAT), which are noisy overall. Nevertheless, this average value is below the expected performance of MATISSE of 5\degr\ \citep{matisse4}. Since FU~Ori's disk is marginally resolved in two baselines, we refrain from interpreting this closure phase signal as anything but a quasi-resolved point source.

The $N$-band closure phase signal is also nonzero with some small variations within the band, and it approaches 5\degr\ at about 10~\micron\ at the largest baseline triangle, which includes the longest UT baseline (green line, top row in Fig.~\ref{fig:Nplots}). This nonzero closure phase signal could point at an asymmetric brightness distribution. As we show further on, a disk inclined in our line of sight can reproduce such a signal (Sect.~\ref{sec:discuss_models}).

\section{Geometric sizes}\label{sec:disksizes}

\begin{figure}
        \centering
                \includegraphics[width=\columnwidth]{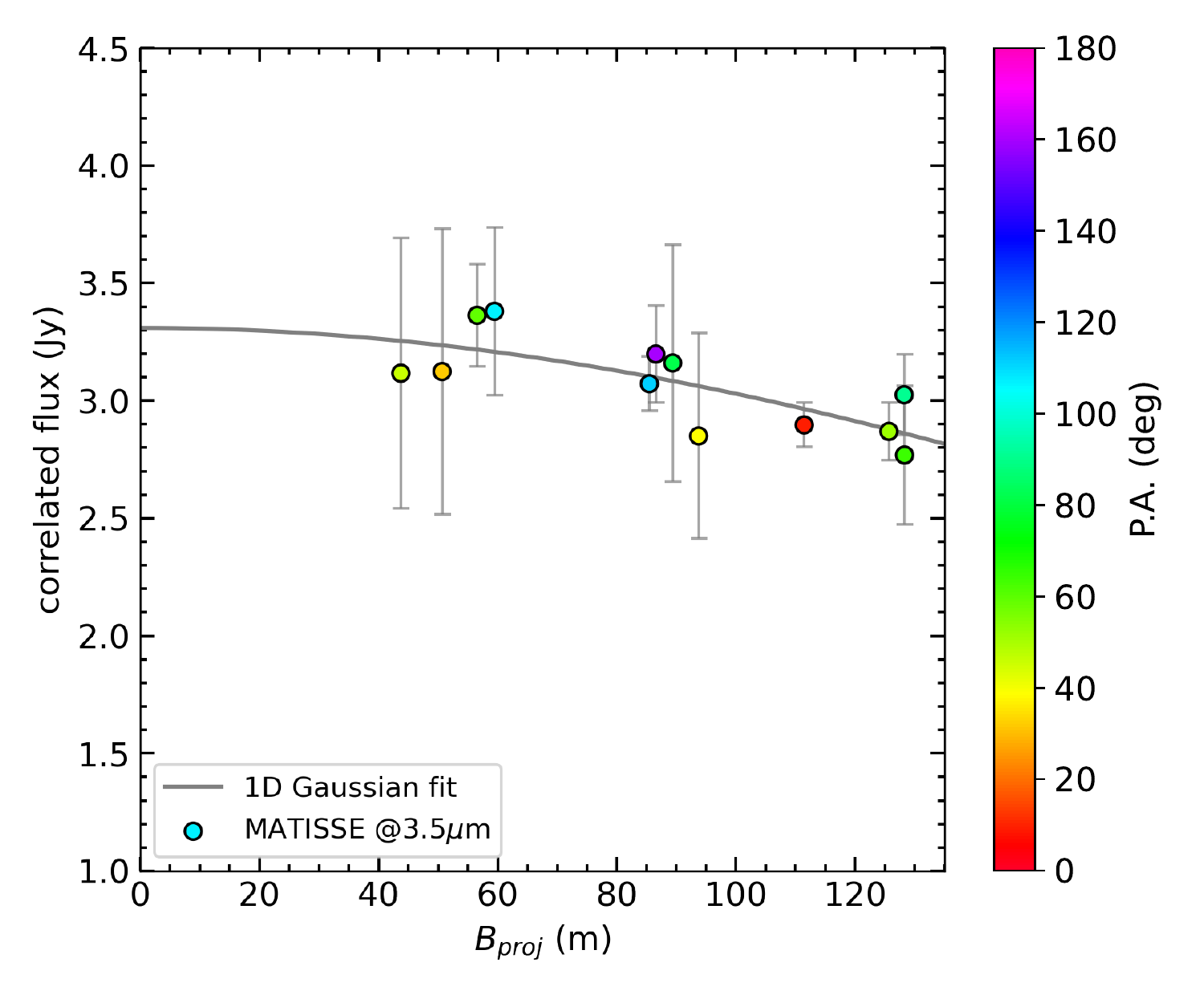}
        \caption{Gaussian (1D) model fit compared with the epoch 4 and 5 correlated fluxes at 3.5\micron.}
        \label{fig:gfits}
\end{figure}

To estimate the geometric properties of the accretion disk from the MATISSE observations, simple centro-symmetric models were fitted to the interferometric data to obtain the angular size and orientation of the detected structure. The Gaussian full width at half maximum (FWHM) for the $L$-band data gave a ``marginally resolved'' angular size of $1.3\pm0.1$~mas at 3.5~\micron, which suggests an apparent size of about 0.5~au for the adopted {\it Gaia} distance (Fig.~\ref{fig:gfits}). 

For the the $N$ band, we opted to model the disk with 2D elliptical Gaussian functions following the method of \citet{2021A&A...647A..56V}. The best-fit to the data is provided with a Gaussian FWHM of $5.3\pm1.4$ mas at 10.5~\micron , a PA of $15\pm25$\degr\ for the minor axis, and an inclination, $i$, of $55\pm15$\degr .

As mentioned in Sect.~\ref{sec:V2}, FU Ori's disk is marginally resolved at the longest baselines in the $L$ band, while in the $N$ band MATISSE was able to resolve a portion of the dusty, passive disk. Therefore, from the geometric sizes estimated above, we can place upper limits on the apparent outer radii of the hot, accretion disk of $\leq 0.3$~au in $L$ and 1~au in $N$.

To assist the reader, throughout the text we accept the conventional orientations for the minor axis PA and disk inclination angle. That is, the PA is measured east-of-north starting at zero degrees north, while $i=90$\degr\ in our line of sight describes an edge-on structure, and a pole-on structure is found at $i=0$\degr. As such, wherever necessary for the context of this work, literature values were adapted to fit these conventions.

\section{Analytical and radiative transfer modeling}\label{sec:simulations}

To further constrain other disk parameters, such as accretion rate, flaring, density distribution, grain sizes, and chemical composition, we opted to apply analytical and radiative transfer models. A description of this process follows below.

In this analysis, we explore two different disk orientations: one with the values derived by MATISSE observations as mentioned above (Sect.\ref{sec:disksizes}), and the other following the geometric properties derived by the Atacama Large Millimeter/submillimeter Array (ALMA) observations \citep{2020ApJ...889...59P}, that is, a disk with PA$=43.6^\circ(\pm1.7)$ and an inclination $i=37.7^\circ(\pm0.8)$. We used both orientations to model both the inner (hot; >1000~K) region and the dusty (passive, cooler) region of the accretion disk.

\subsection{A steady-state accretion disk model}\label{sec:inner}

The majority of the circumstellar emission at shorter wavelengths ($\lambda\leq4\rm\mu m$) originates from the hot, inner accretion disk (i.e., within a radius of 1~au). In our first attempt to model this inner disk, we adopt a model of an optically thick but geometrically thin, viscous accretion disk, where the accretion rate remains constant irrespective of radial distance. 

We follow an approach similar to \citet{2008ApJ...684.1281Z}, although here the synthetic SED is calculated by integrating black-body emission in concentric annuli between the disk's inner radius $R_{\rm in}$, and outer radius $R_{\rm out}$. Therefore, we focus on reproducing the broadband emission, unlike \citet{2008ApJ...684.1281Z} that could simulate spectroscopic features. The disk temperature profile $T(r)$ of this analytical model follows \citet{1996ARA&A..34..207H}:
\begin{equation} 
T (r) = \left[ \frac{3GM\dot{M}}{8\pi R^3 \sigma} \left( 1 - \sqrt{\frac{R_*}{r}} \right) \right]^{1/4},
\end{equation}
where $r$ is the distance from the star, $R_*$ is the stellar radius, $M$ is the stellar mass, $\dot{M}$ is the accretion rate, and $G,\sigma$ are the well-known gravitational and Stefan-Boltzmann constants.

We constrain the model SED based on the contemporary photometry up to 3~\micron\ (Sect.~\ref{sec:phot}, Table~\ref{tab:sed}) corrected for interstellar reddening (Sect.~\ref{sec:extinction}). The initial parameters for the steady-state disk model, that is, the radial extent of the hot inner accretion disk ($R_{\rm in},\,R_{\rm out}$) and the accretion rate ($M\dot{M}$), were taken from \citet{2007ApJ...669..483Z} and \citet{2008ApJ...684.1281Z}. However, since the SED, the reddening, and the distance are now fixed to contemporary values, we found that the aforementioned model parameters had to be adjusted. As such, we opted for a range of $R_{\rm out}$ at 0.2, 0.3, 0.6, 0.8, and 1.0~au, and adjusted the accretion rate ($1\times 10^{-5} \leq M\dot{M} [M_\odot^2/yr]\leq 3\times 10^{-5}$) to obtain an adequate fit to the SED. The best results were achieved for outer radii at 0.3 and 0.6~au. 

Although this simple, analytical approach provided a good description of the hot, inner disk of FU~Ori, and could adequately fit the SED up to 2.5~\micron, such a model cannot simulate the entire accretion disk, and especially its dusty component. We expect that the dusty passive disk would dominate at the mid-infrared and at longer wavelengths, while having a minor contribution in the near-infrared.

We therefore opted to model both the inner and the dusty (passive) accretion disk components with the Monte Carlo radiative transfer simulation tool {\sc radmc-3d} \citep{radmc3d}, which can treat different geometries and dust compositions. The radiative transfer model is described in the next section. 


\subsection{Radiative transfer model}\label{sec:outer}

We modeled FU Ori's disk by employing the  Monte Carlo radiative transfer simulation tool {\sc radmc-3d} \citep{radmc3d}, which can treat different geometries and dust compositions. For each model setup, including the user-specified spatial distribution of circumstellar material, and the knowledge of the optical properties of the material, {\sc radmc-3d} could solve the radiative equilibrium problem and find out the radiative equilibrium temperature distribution. Multiwavelength brightness distribution could then be computed, using the information of density and temperature distribution and opacity properties.

We adopt a model setup similar to \citet{2008ApJ...684.1281Z} and \citet{2020ApJ...889...59P}, including two components, that is, an inner accretion disk and an outer passive disk. The input parameters for the inner accretion disk were initially based on the best-fit results provided by the analytical model (see Sect.~\ref{sec:inner}), but we allow for minor adjusting.

The spatial distribution of dust and the accretion heating source is described in the following. For each disk component, we assume a power-law density distribution. The disk density profile is described as
\begin{equation}
    \rho = \frac{\varSigma (r,\phi)}{H\sqrt{2\pi}}\exp{\left( -\frac{z^2}{2H^2} \right)}\hspace{5pt} ,
\end{equation}
where $z$ is the height above the disk mid-plane, $H$ is the scale height, and $\varSigma$ is the surface density profile. The last two parameters can be described by power-law relations as 
\begin{equation}
    \frac{H(r)}{r} = h_{\rm ref}\left( \frac{r}{R_{\rm ref}} \right)^{q} ,
\end{equation}
where $q$ is the disk flaring parameter and $h_{\rm ref}$ is the ratio of the scale height at the reference radius $R_{\rm ref}$, while 
\begin{equation}
    \varSigma(r) = \varSigma_{\rm ref} \left(\frac{r}{R_{\rm ref}} \right)^{-p} ,
\end{equation}
where $\Sigma_{\rm ref}$ is the surface density at the disk's inner radius and $p$ is a power-law exponent. We set $R_{\rm ref}$ equal to the radius where the inner and outer disk are separated.

To account for the accretion heating in the inner disk, we put the heating source in the mid-plane with power per area of
\begin{equation}
    P = \frac{3GM\dot{M}}{4{\pi}r^3} \left[1 - \left( \frac{R_*}{r} \right)^{1/2} \right], \,\mathrm{where}\,\, R_\mathrm{in,1}<r<R_\mathrm{out,1} .
\end{equation}
\subsubsection{Dust composition}\label{sec:dust}
In the following we explain our choice of the dust composition of the passive disk. The shape of FU Ori's silicate feature indicates dust grain growth and deficiency of submicron grains. \citet{2006ApJ...648..472Q} fitted the silicate feature using a mixture composed mainly of amorphous silicate with olivine and pyroxene stoichiometry, with grain sizes from 0.1 to 6 $\mu$m. However, their study does not exclude the existence of larger grains (${>}10~\mu$m). Based on this, we adopt a mixture of two dust components, which we named ``small silicate'' and ``larger.''
The small silicate component, in turn, is the same mixture as described in the Table~4 of \citet{2006ApJ...648..472Q}.
The larger component is a mixture of carbonaceous material and ``astronomical silicate'' \citep{1984ApJ...285...89D} with a mass ratio of C:Si=1:2, and a power-law grain size distribution $f(a)\sim a^{-3.5}$ \citep{mrn}, from $a_{\rm min}=10$~\micron\ to $a_{\rm max}=1000$~\micron.
The mass ratio was set to this value because recent studies generally indicate that there is more silicate than carbon in protoplanetary disks \citep[e.g.,][]{2021A&A...649A..84H}.
The small silicate and larger dust were then mixed with a mass ratio of $f_\mathrm{small}$:(1-$f_\mathrm{small}$), where $f_\mathrm{small}$ is a free parameter and represents the percentage of small grains in the mixture. It is constrained by the strength of the silicate feature.
The dust opacities were derived using the {\sc OpacityTool} software \citep{1981ApOpt..20.3657T,2005A&A...432..909M,2016A&A...586A.103W},
which is based on the distribution of hollow spheres (DHS) theory \citep{2005A&A...432..909M}.
In the DHS theory, for a ensemble of dust grains with irregular shape,
the optical effects of grain shape are represented by a single shape parameter $f_\mathrm{max}$.
{\sc OpacityTool} can compute the optical properties for a given dust mixture, if the dust grain size and shape distribution, as well as the complex refractive index of the dust species as a function of wavelength are known. The complex refractive index data were collected from the literature
(%
\citealt{1995A&A...300..503D}, amorphous silicate;
\citealt{1984ApJ...285...89D}, \citealt{1993ApJ...402..441L}, \citealt{2001ApJ...548..296W}, astronomical silicate;
\citealt{1998A&A...332..291J}, carbonaceous material%
).
We adopt a grain shape parameter of $f_{\rm max}=0.8$ following \citet{2016A&A...586A.103W}.

For the inner accretion disk, the opacity ($\tau$) is dominated by gas free-free absorption, which is temperature-dependent, and therefore the radiative transfer process cannot be accurately simulated with {\sc radmc-3d}. However, this does not significantly affect the optical and near-infrared continuum, as long as the gas density is high enough to ensure $\tau\gg1$. As such, and for simplicity, we represent the inner disk with an artificial gray material with $\kappa_{\rm abs} = 1$~cm$^2$g$^{-1}$ and $\kappa_{\rm sca} = 0$~cm$^2$g$^{-1}$ at any wavelength. 

\subsubsection{Disk geometry}

As described above (Sect.~\ref{sec:inner}), the inner disk was modeled as a geometrically thin one. This presumes that its scale height is quite small compared with the dusty component. We first attempted to model the inner component as a ``thick slab'' geometry in {\sc radmc-3d}, that is, with scale height $h\neq0$ and flaring index $q_1=0$. However, this could not reproduce the optical to near-infrared SED well, and we opted for a slightly flared inner disk.

Scattered-light images of FU~Ori suggested a disk size of ${\sim}80~$au \citep{2020ApJ...888....7L}. However, at radio wavelengths the disk appears slightly more compact. \cite{2020ApJ...889...59P} modeled the $1.3~$mm continuum image using a disk with outer radius of 100~au,
but with a characteristic radius of just $11.3$~au, outside of which the surface density drops quickly. Considering this information, we also adopt in our model an outer radius of 100~au, but use a steep power-law for the surface density $\Sigma\propto r^{-2}$, so that the disk is optically thick at $1.3$~mm only in the inner ${\sim}10$~au region. We also tested models with smaller (60 au) and larger (150~au) outer radii, but these failed to reproduce the mid-infrared to submillimeter emission in the SED.

Under the framework described above, we adjusted the model parameters to fit the observations, including the contemporary SED, the correlated fluxes, and the closure phases in the MATISSE observations. As mentioned in the previous section, we found that analytical models with inner disk outer radii at 0.3 and 0.6 au, provided an adequate fit to the SED for the two disk orientations. However, when these were used as start-up parameters for the {\sc radmc-3d} model, models at 0.6~au produced more flux below 5\micron, and models at 0.3~au provided better fits once the accretion rate was slightly reduced (order of $1\times10^{-5}\,M_{\sun}^2$~yr$^{-1}$).

Table~\ref{tab:parameter:RT} lists the parameters for the two disk orientations. In general, we have not optimized our parametric search (e.g., with a $\chi^2$ minimization or Markov chain Monte Carlo process) but selected parameters based on visual inspections of the data versus model comparisons. Therefore, we do not provide errors (or a range of values) for each parameter; we only present our initial parameter-space search (cf., Sect.~\ref{sec:inner}).  Certain parameters, such as the disk's orientation and the inner component's surface density, are fixed (boldface in Table~\ref{tab:parameter:RT}), while some of the outer component's parameters are inherited from the inner one (e.g., $R_{\rm in,2}\equiv R_{\rm out,1}$).

\begin{figure}[htb]
    \centering
    \includegraphics[width=10cm]{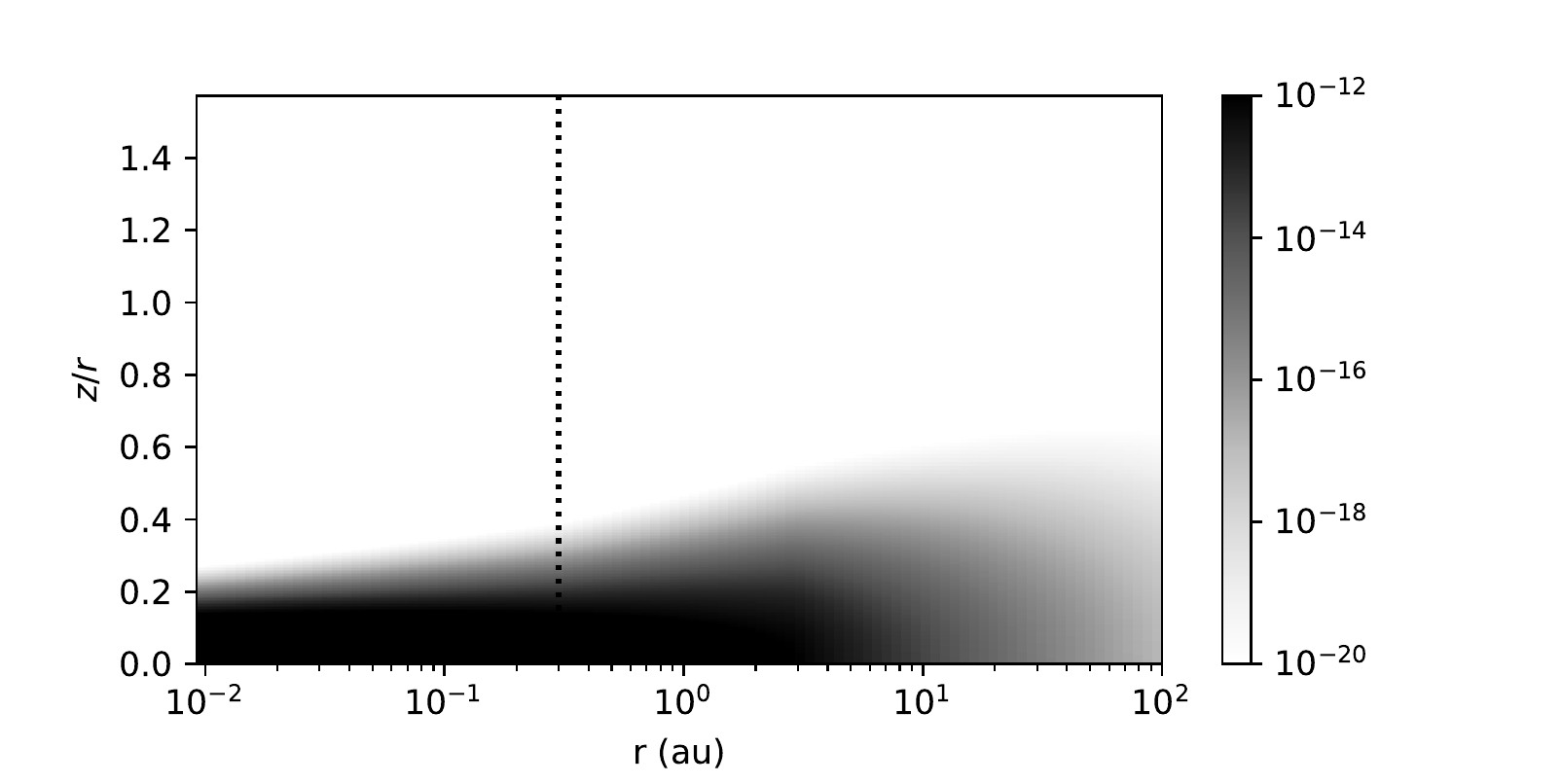}\\
    \includegraphics[width=10cm]{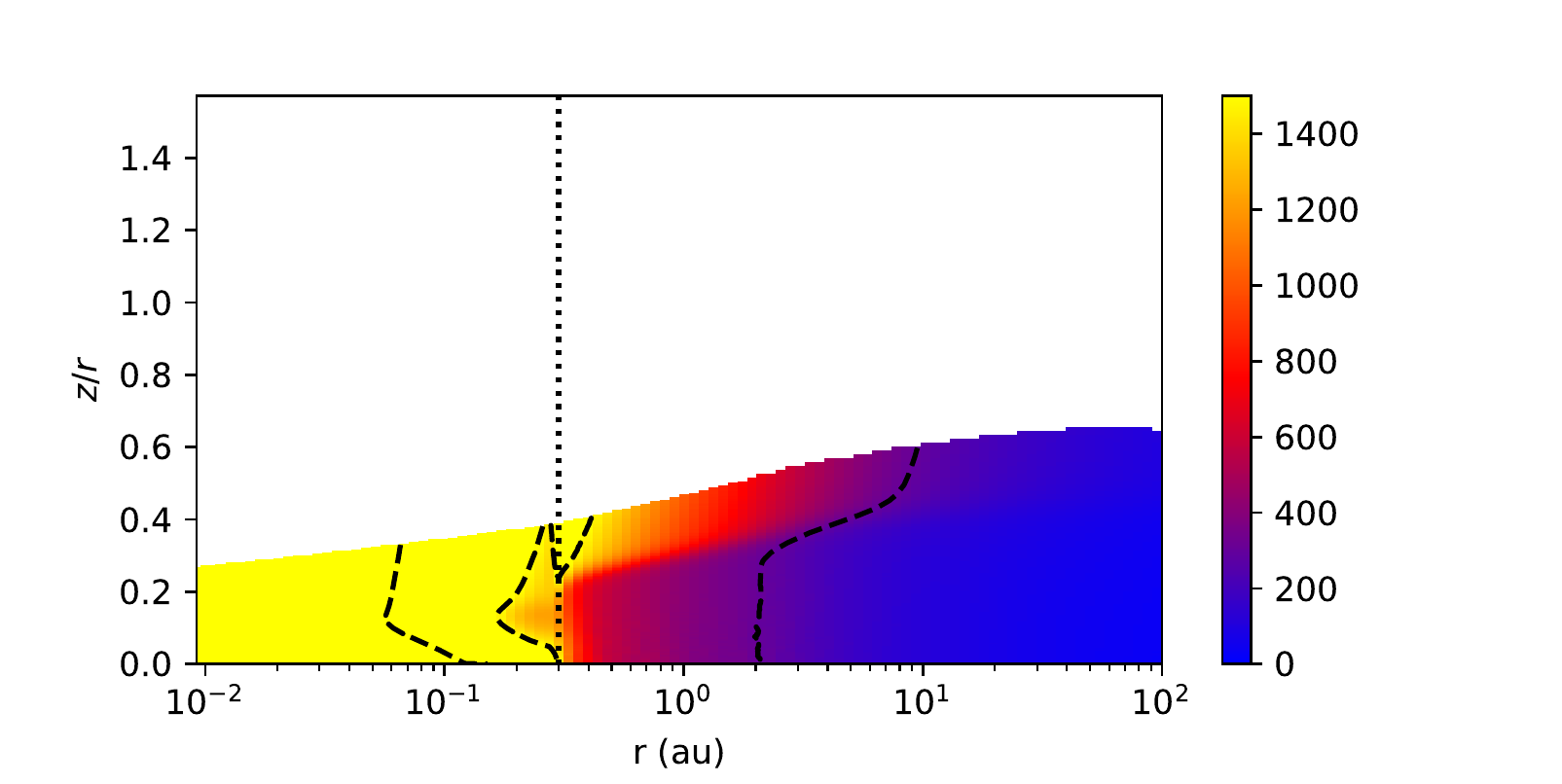}
    \caption{Disk structure in the best-fit \textsc{radmc-3d} radiative transfer model with ALMA orientation.
    \textbf{Top:} Disk density profile. The colorbar indicates units of g cm$^{-3}$. \textbf{Bottom:} Disk temperature profile. The colorbar is in units of kelvin. The vertical dashed lines mark the edge between the inner accretion disk and outer passive disk (0.3 au).
    In the temperature distribution, contour lines for temperatures of $3000$~K, $1500~$K, and $300$~K are plotted.
    }
    \label{fig:DiskStructure}
\end{figure}

\begin{figure}[htb]
    \centering
    \includegraphics[width=\columnwidth, angle=180]{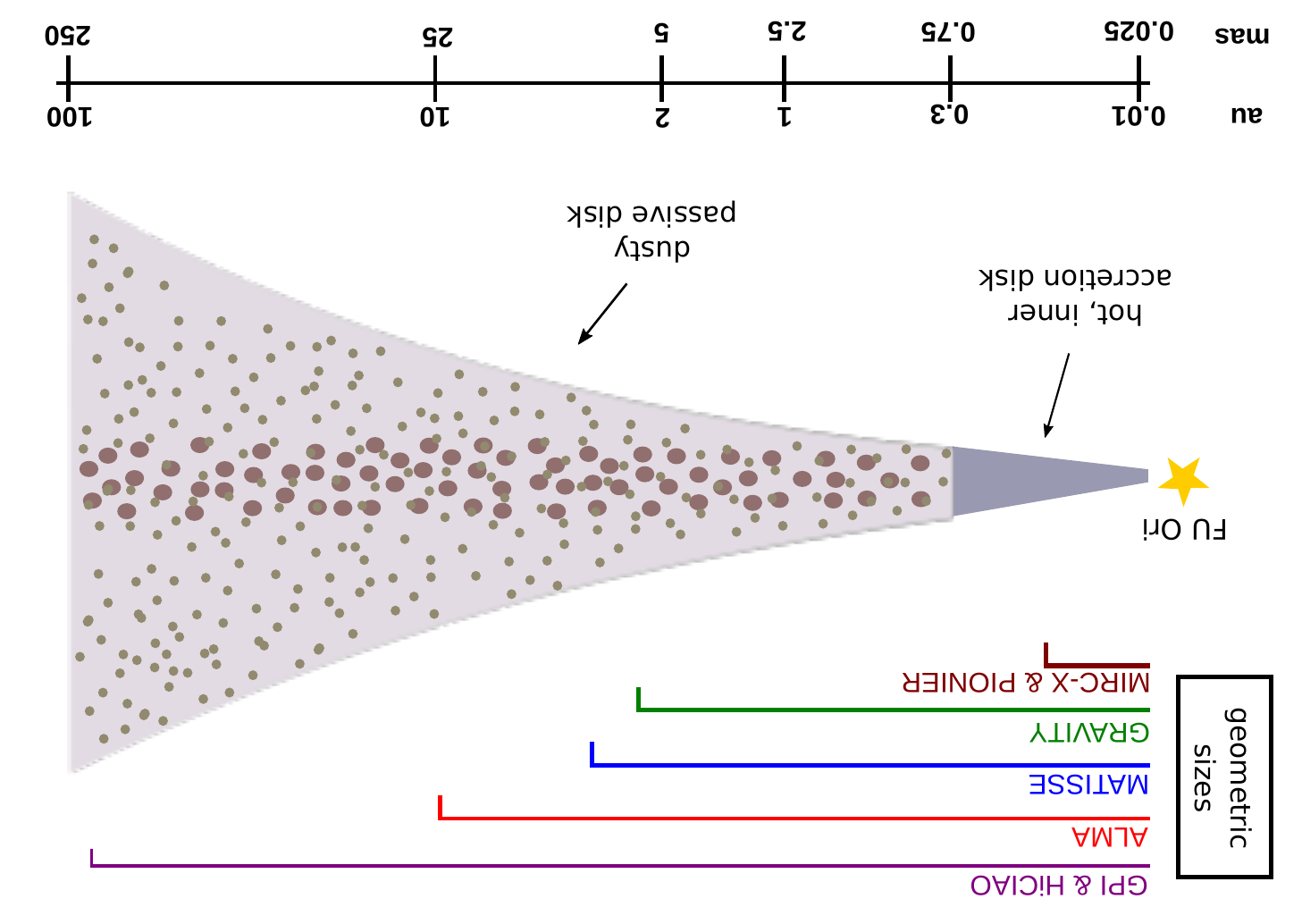}
    \caption{Rudimentary sketch of the FU Orionis accretion disk, indicating the inner, hot region (dark gray area; component 1) and the dusty, passive disk (light gray area; component 2), which at 0.3 au are separated. In this work, we presume that the inner disk is optically thick and devoid of dust, although it could be surrounded by a halo of gaseous material. The dusty disk is flared, with the bulk of large grains settling in the mid-plane. We roughly indicate the extent of the regions (geometric sizes and/or direct images) that have been detected by MATISSE and by other high-angular-resolution instruments, such as MIRC-X and the Precision Integrated-Optics Near-infrared Imaging ExpeRiment (PIONIER) ($H$), GRAVITY ($K$), ALMA (1.3mm continuum), and the GPI and High-Contrast Coronographic Imager for Adaptive Optics (HiCIAO) (scattered light; $J,H$). The physical scales are given here in astronomical units, and they are converted to angular sizes, with the {\it Gaia} EDR3 distance to FU Ori adopted. Other than that, this drawing is not to scale.}
    \label{fig:sketch}
\end{figure}

We illustrate in Fig.~\ref{fig:DiskStructure} the disk structure (i.e., disk density and temperature profiles) of the {\sc radmc-3d} model with the ALMA orientation, noting that the model with the MATISSE orientation has a similar structure. We also provide a rudimentary sketch of FU Ori's disk in Fig.~\ref{fig:sketch}. Here, we indicate the regions of the disk that have been detected by MATISSE and by other high-angular-resolution instruments. These are marked based either on direct images of the disk -- for example from ALMA and the Gemini Planet Imager (GPI) -- or on geometric sizes -- from the Michigan InfraRed Combiner-eXeter (MIRC-X) of the Center for High Angular Resolution Astronomy (CHARA), GRAVITY/VLTI, and MATISSE/VLTI.

\begin{figure}[htb]
    \centering
        \includegraphics[width=\columnwidth]{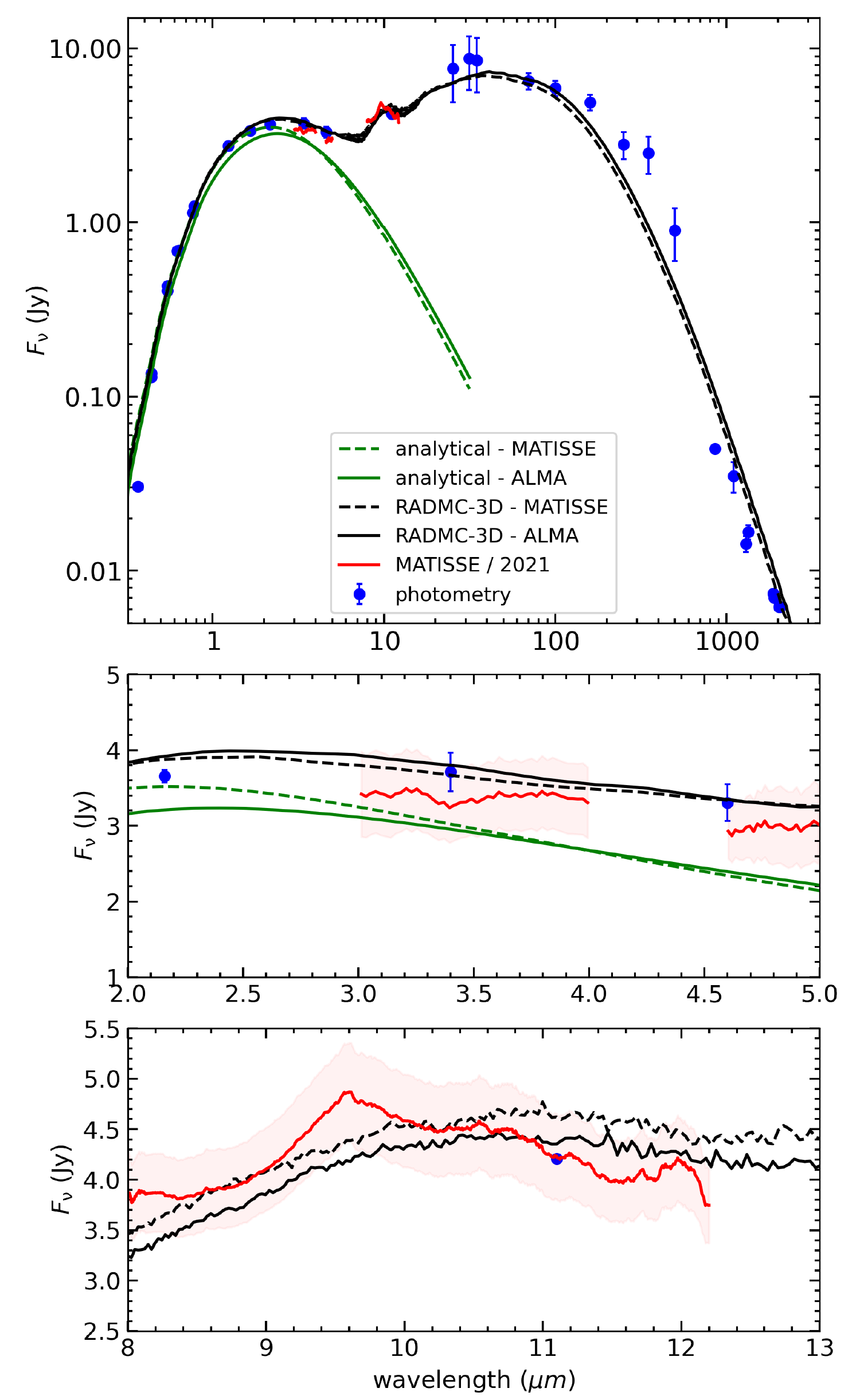}    
    \caption{Spectral energy distribution comparing the contemporary photometry (blue circles; Table~\ref{tab:sed}) and the MATISSE spectrum (red; epoch 4) against the analytical (green) and {\sc radmc-3d} (black) models with the two disk orientations: ALMA (dashed lines) and MATISSE (solid lines). The bottom two panels are enlargements of the $L$- and $N$-band regions. The shaded regions represent the flux uncertainties of the MATISSE spectra, while the 9.4 - 9.9\micron\ spectral region is heavily affected by the terrestrial atmosphere.}
    \label{fig:sed}
\end{figure}

The synthetic SEDs of the models described in Table~\ref{tab:parameter:RT} are shown in Fig.~\ref{fig:sed}. The synthetic correlated fluxes at 3.5\micron\ and at 10\micron\ are compared with observations in Figs.~\ref{fig:models-L-fcorr} and \ref{fig:models-N-fcorr}, respectively, while the $N$-band synthetic closure phases are shown in Fig.~\ref{fig:models-N-cp}. Furthermore, we show the synthetic visibilities and correlated fluxes in the entire $N$ band in Figs.~\ref{fig:extra_vis2} and \ref{fig:extra_fcorr}.

\begin{table}[htbp]
    \caption{{\sc radmc-3d} model parameters for two disk orientations. They correspond to the models shown in Figs.~\ref{fig:DiskStructure}, \ref{fig:sed}-\ref{fig:models-N-cp}, and \ref{fig:extra_vis2}-\ref{fig:extra_fcorr}. Fixed parameters are marked in boldface font.}
    \label{tab:parameter:RT}
    \centering
    \begin{tabular}{lcc}
    \hline
    Parameter & ALMA & MATISSE  \\
    \hline\hline
    {\bf PA} (deg) & {\bf 43.6} & {\bf 15 }\\
    $\bm i$ (deg) & {\bf 37.7} & {\bf 55} \\ \hline
    \multicolumn{3}{c}{\it Inner disk (Component 1)} \\
    $M\dot{M}$ ($M_\sun^2/\rm{yr}$)  & $0.85\times10^{-5} $ & $1.64\times 10^{-5} $ \\
    $R_{\rm in,1}\,(R_\sun)$  & 1.98 & 2.33 \\
    $R_{\rm out,1}$ (au) & 0.30 & 0.30 \\
    $\bm \Sigma_{\rm in,1}$ ($\,\rm{g/cm^2}$) & {\bf 10.0 } & {\bf 10.0 } \\
    $\bm p_1$ & {\bf 0 }     & {\bf 0 } \\
    $h_{\rm ref,1}$  & 0.06   & 0.06  \\
    $q_1$    & 0.13 & 0.13 \\
    \hline
    \multicolumn{3}{c}{\it Outer disk (Component 2)} \\
    $R_{\rm in,2}$ (au)  & $\equiv R_{\rm out,1}$ & $\equiv R_{\rm out,1}$ \\
    $R_{\rm out,2}$ (au) & 100 &  100 \\
    $\varSigma_{\rm in}$ ($\rm{g/cm^2}$)  & 940 & 904 \\
    $p_2$ & $-2.0$ & $-2.0$ \\
    $h_{\rm ref,2}$  & $\equiv h_{\rm ref,1}$ & $\equiv h_{\rm ref,1}$ \\
    $q_2$  & 0.18 & 0.18 \\
    $f_\mathrm{small}$  &   0.10 & 0.05 \\
    \hline
    \end{tabular}
    \tablefoot{The dust mass derived from these models is $2.4\times10^{-4}$~$M_\sun$ and $2.3\times10^{-4}$~$M_\sun$ for the ALMA and MATISSE orientations, respectively.}
\end{table}


\section{Discussion}\label{sec:discuss}

\subsection{The current SED of FU Orionis}\label{sec:sed}
FU~Orionis has been steadily fading since it reached peak brightness in 1937. \citet{2000ApJ...531.1028K} estimated a decline rate of 0.015 mag yr$^{-1}$ in the $B$ band. This would suggest a drop in magnitude of 1.3 mag since 1937 (when the peak brightness was $\sim$9.8 mag and assuming $m_{\rm pg}$ is equivalent to $B$) placing the current brightness at 11.1 mag in $B$.

Following this, we opted to complement our interferometric observations with contemporaneous photometric observations (Table~\ref{tab:sed}; Sect.~\ref{sec:phot}), which enables us to model the current properties of FU~Ori's accretion disk. Our photometric measurements are in agreement with the expected decline as above, taking into account that FU Ori itself is a short-term variable source \citep[$\Delta V\sim0.035$ mag within 1 day;][]{2000ApJ...531.1028K}.

Whilst comparing the SED from our observations to literature studies, we note a shift, a decline in brightness, in the near-infrared as well. For example, the system has faded by approximately 0.5~mag in the $K$ band within the last 20 years. Since the photometry indirectly influences the parametric search in the analytical and radiative transfer simulations (Fig.~\ref{fig:sed}, upper panel; Sect.~\ref{sec:simulations}), it is expected that our model results will differ from previous works \citep[e.g.,][]{2007ApJ...669..483Z, 2020ApJ...889...59P, 2021A&A...646A.102L}. We would therefore caution the reader in comparing any new disk simulations to photometric measurements from the past literature. A further analysis of the past evolution will be presented in a forthcoming paper (Lykou et al., in preparation). 



\begin{figure}[htbp]
\centering
    \includegraphics[width=8.5cm]{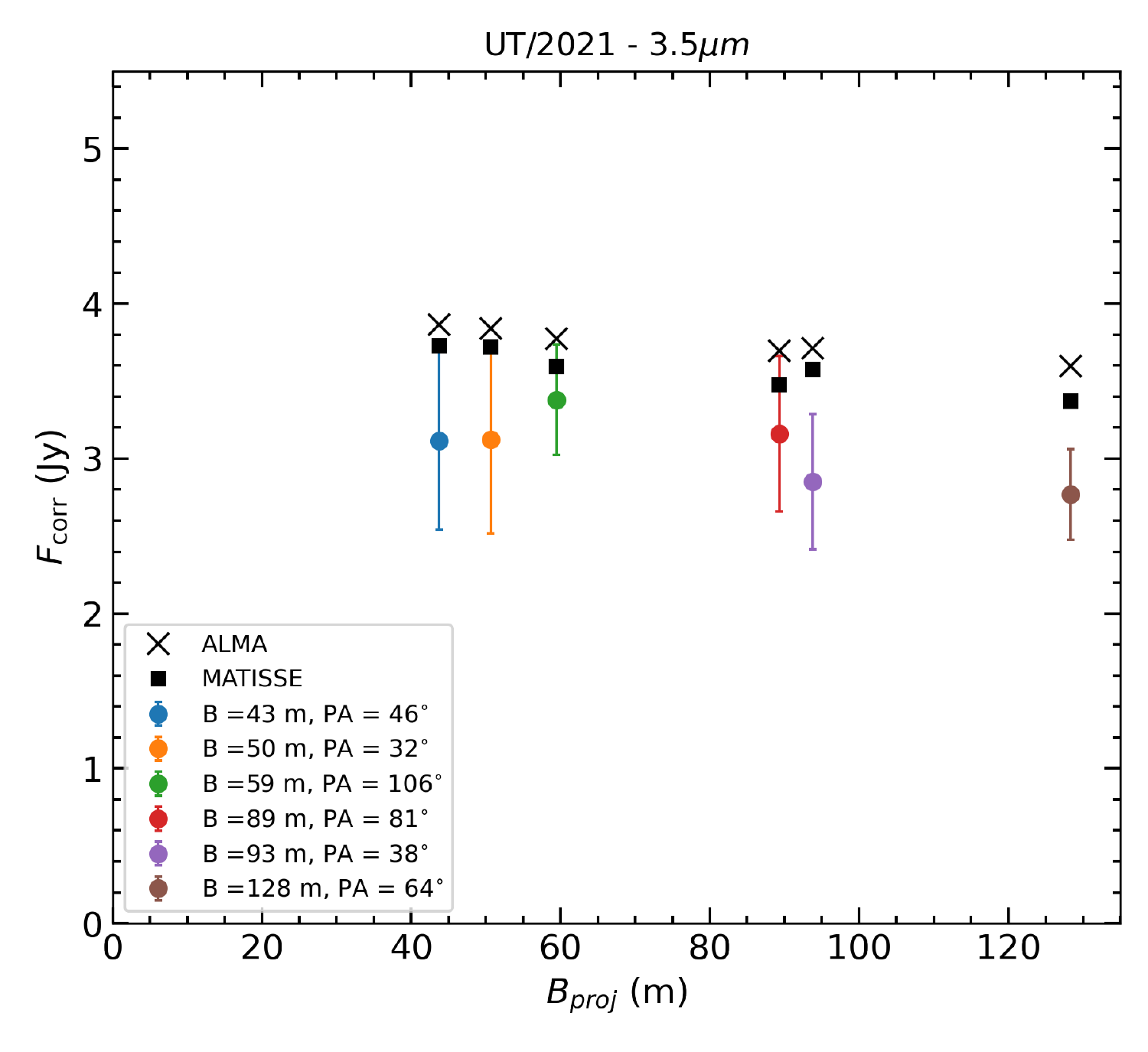}\\
    \includegraphics[width=8.5cm]{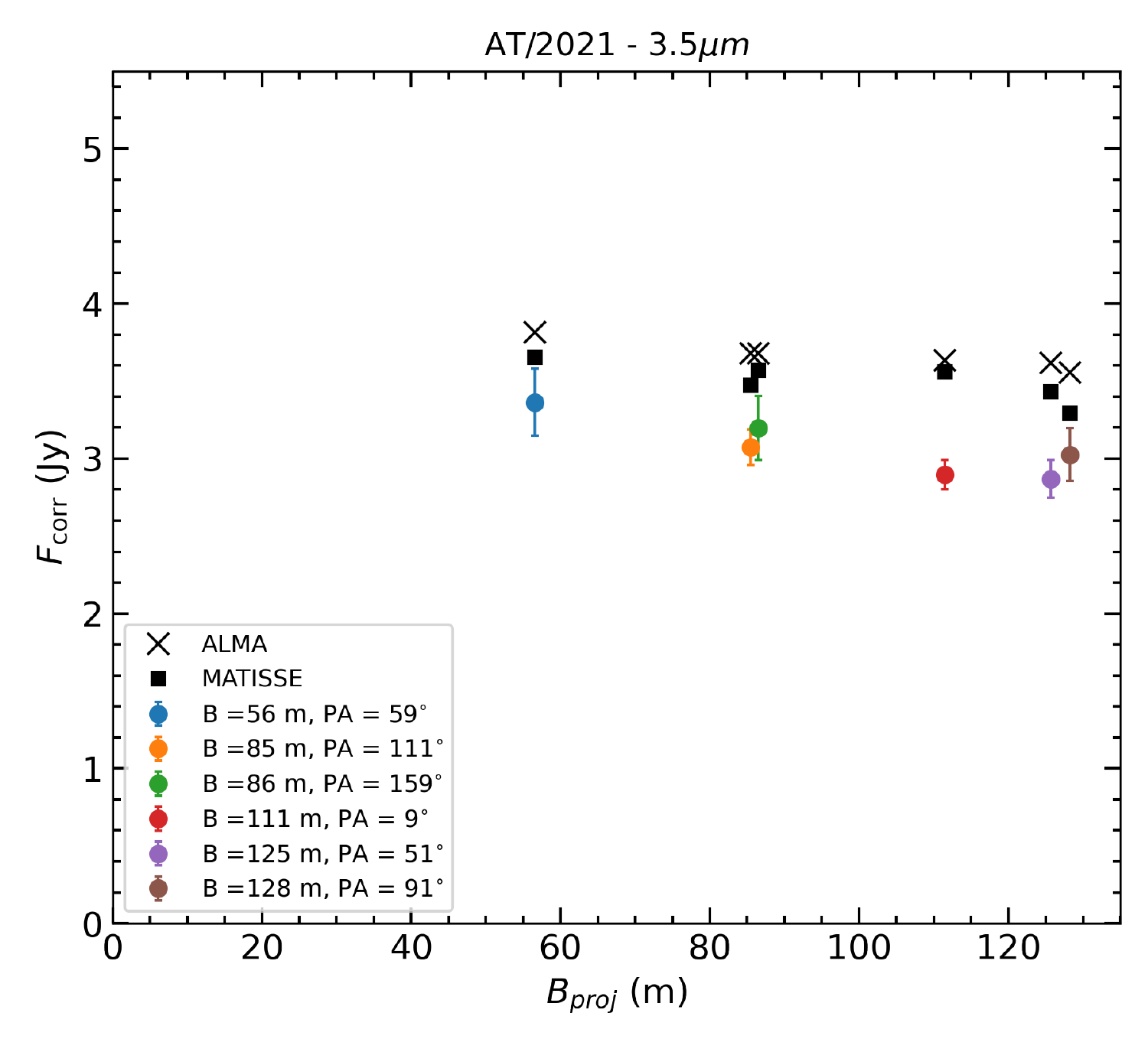}
        \caption{Synthetic correlated fluxes at 3.5~\micron\ from the best-fit {\sc radmc-3d} models of Table~\protect\ref{tab:parameter:RT} (ALMA: crosses; MATISSE: squares) compared with MATISSE data, color-coded per respective baseline: UTs (top panel) and ATs (bottom panel).
        }
        \label{fig:models-L-fcorr}
\end{figure}

\begin{figure}[htbp]
    \centering
        \includegraphics[width=8.5cm]{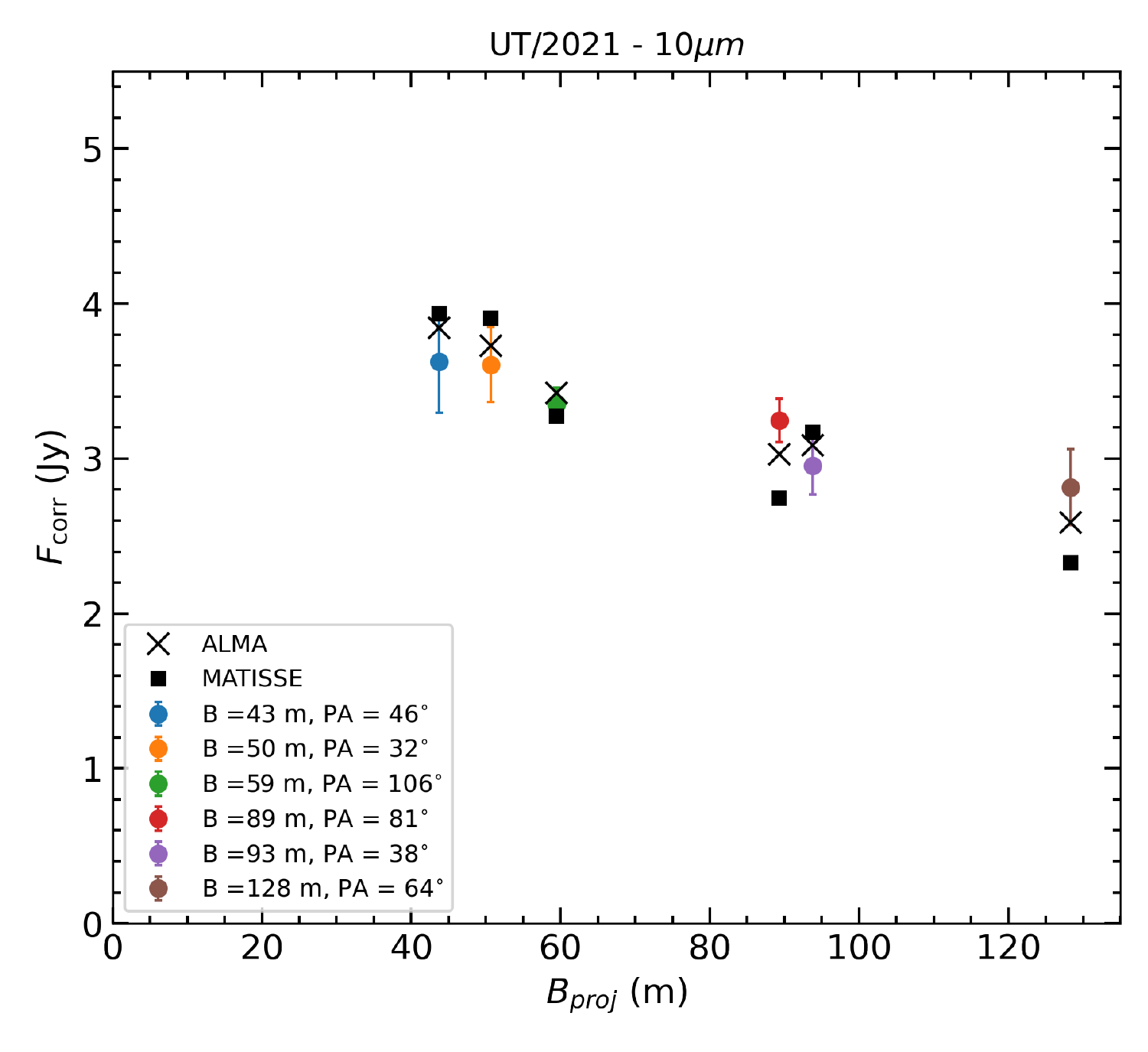}
    \caption{As in Fig.~\ref{fig:models-L-fcorr} but for 10~\micron\ and only for the UT (epoch 4) data.}
    \label{fig:models-N-fcorr}
\end{figure}

\begin{figure}[htbp]
    \centering
    \includegraphics[width=8cm]{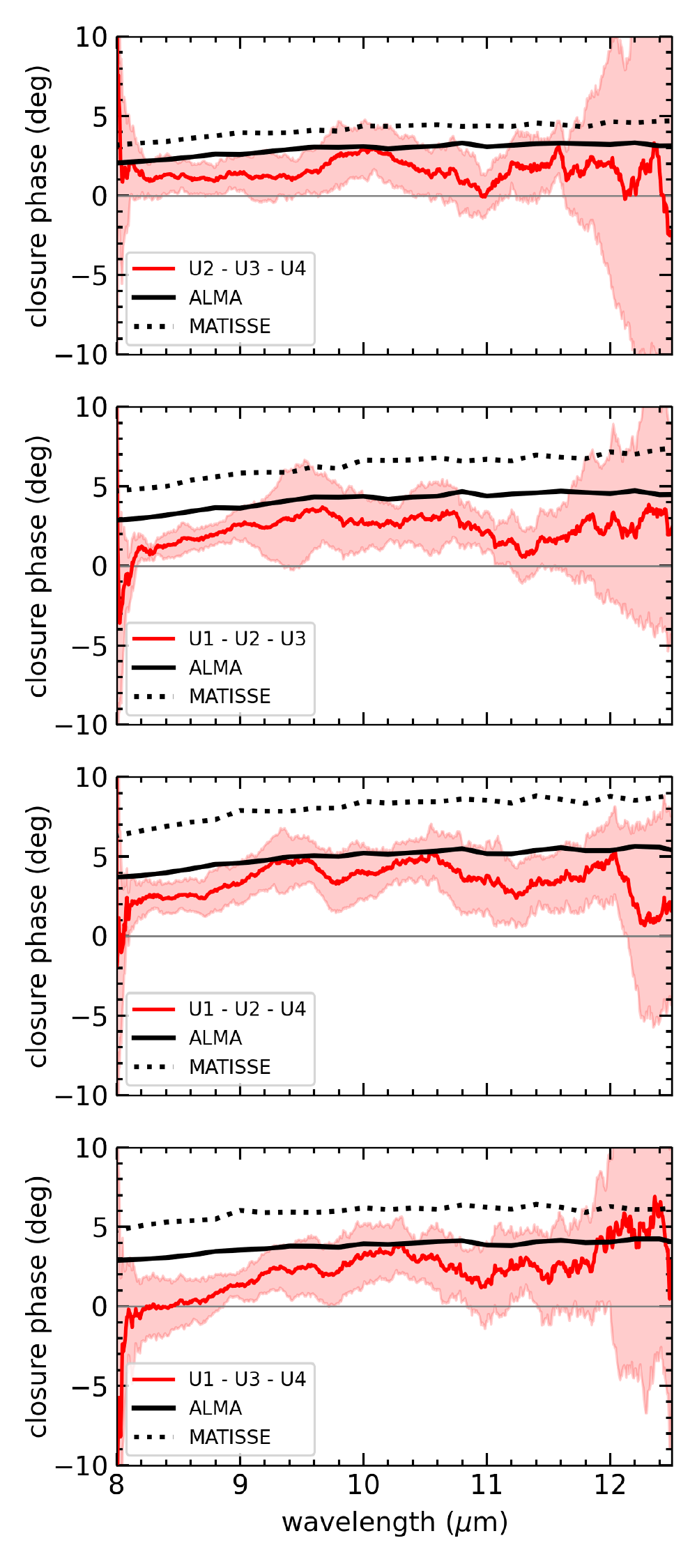}
    \caption{Synthetic $N$-band closure phases from the two {\sc radmc-3d} disk models (ALMA model as a solid black line, and MATISSE model as a dotted line) compared with the MATISSE data (red; epoch 4).}
    \label{fig:models-N-cp}
\end{figure}


\subsection{Radiative transfer model results}\label{sec:discuss_models}

Our approach to simulating FU~Ori's disk with different models and the subsequent ``best-fit'' parameters\footnote{These models are not unique, given the large number of free parameters, but we note them here as ``best-fitting.'' Future observations may provide better constraints for these models.} was presented in Sect.~\ref{sec:simulations}. An initial parametric search was performed with a steady-state accretion disk model for the inner hot disk. This provided an adequate fit to the optical and near-infrared broadband photometry, but the models under-estimate the flux beyond 3~\micron. We later used these parameters from the analytical approach, as a starting model for the {\sc radmc-3d} simulations, where the (featureless) inner, hot accretion disk acts like an illuminating source for the dusty disk. These radiative transfer simulations required further adjustment of initial parameters, such as the accretion disk's outer radius and mass accretion rate, to achieve an adequate fit for the entire SED and interferometric data. In the following, we focus on the {\sc radmc-3d} model results.

Despite the uncertainty in the absolute photometric calibration of the MATISSE $N$-band spectrum, we find that both models can adequately fit the MATISSE $L$-, $M$-, and $N$-band spectra within the uncertainties (Fig.~\ref{fig:sed}). This includes the flat-topped silicate feature that is evident in all FU Ori spectra (Fig.~\ref{fig:LMNflux}).

A first major result of the radiative transfer analysis is that the inner disk's outer radius is found to be smaller (0.3~au) than literature values (e.g., \citealt{2007ApJ...669..483Z} : 0.5 - 1 au; \citealt{2021A&A...646A.102L} : $0.74\pm0.35$ au). This appears to agree with the MATISSE results where we find that the $L$-band emitting region ought to be smaller than 0.5~au in diameter. \citet{2021ApJ...923..270L} hypothesized that the hot (inner) disk region may be cooling down, which may be supported by a temporal decline in the SED, and could lead to a shrinkage of that region. When opting for larger (outer) radii, we found that the model flux was much higher in the 2 - 6~\micron\ and 10 - 30~\micron\ regions irrespective of the disk orientation. The ``best-fit'' models shown here (Fig.~\ref{fig:sed}) overestimate the flux by approximately 10\% in the 2 - 6~\micron\ region. However, the fits fall within the uncertainties of the MATISSE photometry. 

Figure~\ref{fig:models-L-fcorr} shows a comparison of the synthetic correlated fluxes at 3.5~\micron\ from the two disk orientations (ALMA as crosses, and MATISSE as squares) with the MATISSE data from the UTs and the ATs. At first glance, there is no clear distinction between the ALMA and MATISSE geometries. On the other hand, both disk models overestimate the correlated fluxes. We attribute this to the overestimation of the synthetic total flux in the 2 - 6~\micron\ region as discussed above.

By fixing the outer radius at 0.3~au for both disk orientations, we first had to adjust the accretion rate\footnote{Adopting a stellar mass of 0.6~$M_\sun$ \citep{2020ApJ...889...59P}. } for each model to $1.4\times10^{-5}$ and $2.7\times10^{-5}$~$M_\sun$ yr$^{-1}$ for the ALMA and MATISSE orientations, respectively (Table~\ref{tab:parameter:RT}). The remaining parameters were kept similar for both models, except for $f_{\rm small}$, the percentage of small silicate grains in the dust mixture (Sect.~\ref{sec:dust}). We find that by reducing $f_{\rm small}$ from 10\% (as used in the ALMA model) to 5\%, the MATISSE model can also provide an adequate fit to the SED and visibilities.

A distinction between the two disk orientations can be seen in the $N$-band data. It appears that the ALMA orientation provides a better fit to both the correlated fluxes at 10~\micron\ (Fig.\ref{fig:models-N-fcorr}) and the closure phases (Fig.~\ref{fig:models-N-cp}). This suggests that a pole-on geometry is more favorable. This reinforces our argument that the nonzero closure phases (Sect.~\ref{sec:CPs}) can be interpreted by a flared disk inclined to our line of sight.

\subsubsection{Disk mass}

Our models suggest a disk mass (dust) of $2.4\times10^{-4}$~$M_\sun$. This is in line with previous studies \citep{2015ApJ...812..134H}, which placed the dust mass at $2\times10^{-4}$~$M_\sun$ for optically thick dust in the submillimeter regime. On the other hand, our result is a factor of 3 higher than the dust mass predicted by \citet{2020ApJ...889...59P}, and an order of magnitude less than what \citet{2019ApJ...884...97L} derived  from radio observations ($1.8\times10^{-3}$~$M_\sun$).

For a typical gas-to-dust ratio of 100, similar to the interstellar medium value, we obtain a lower limit for the total disk mass (gas and dust) of $\sim 0.02$~$M_\sun$. FU~Ori's disk is therefore still one of the most compact and least massive in its class \citep{2018A&A...612A..54L, 2021ApJS..256...30K}. 

The derived total disk mass is slightly higher than the expected amount of material accreted onto FU~Ori since its eruption, but it is still a small amount. That is, if one assumes that the accretion rate remained constant at $\sim 1\times 10^{-5}$~$M_\sun$\ yr$^{-1}$ (at the order of the accretion rate derived by our models) for the first 85 years since the eruption, then the total accreted mass would have been $8.5\times 10^{-4}$~$M_\sun$. The remaining disk mass, if it is indeed 0.02~$M_\sun$, would be accreted within 2000 years at this rate.

\subsection{Dust mineralogy and radial variations}

Here we explore any potential radial variations in the silicate feature within the disk itself. The {\sc radmc-3d} models predict a small radial dependence of the silicate feature with respect to spatial sampling. That is, correlated spectra from the shortest baselines, which sample larger portions of the disk, indicate stronger silicate emission as opposed to the longest baselines, where the emission diminishes. This is of course expected, since the silicate emission is stronger when integrating over larger areas of the silicate-rich disk. On the other hand, the longest MATISSE baseline is probing a region of the simulated disk that is silicate-free in this case, that is, the inner accretion disk, and hence no silicate emission arises from that area. This hypothesis will be examined in the following section.

\begin{figure}
    \centering
    \includegraphics[width=\columnwidth]{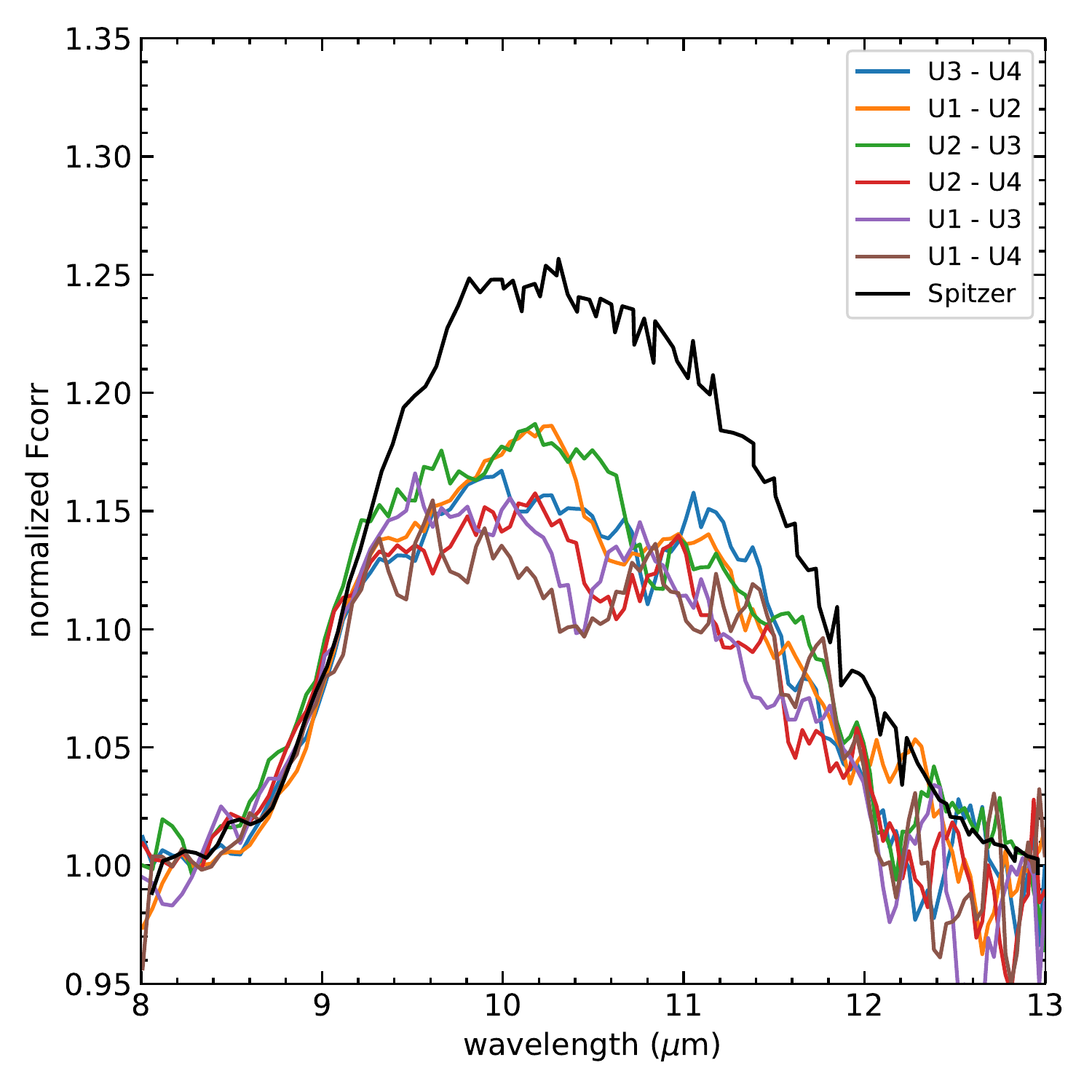}
    \caption{Normalized correlated spectra from MATISSE (color-coded per baseline) against {\it Spitzer} \citep[in black;][]{2006ApJ...648.1099G}. }
    \label{fig:compsilicates}
\end{figure}

We compare the MATISSE/VLTI $N$-band correlated spectra with the {\it Spitzer} spectrum from 2004 \citep{2006ApJ...648.1099G} in Fig.~\ref{fig:compsilicates}. For the sake of clarity, we normalize  all spectra following the method of \citet{2003A&A...400L..21V}. That is, we obtain a linear fit between 8.3 and 12.9~\micron, which is defined as the ``continuum,'' which we then subtract from the spectrum. As last step, we normalize the subtracted spectrum with the mean of the continuum. The final product is scaled to be larger than unity for clarity. Overall, we do not notice any significant differences in the shape of the silicate feature per baseline. In fact, the silicate feature appears to be flat-topped, and thus resembles emission arising from larger-sized grains, such as $\alpha\geq2\,\rm\mu$m \citep[e.g., Fig.2 of][]{2001A&A...375..950B}. 

Unlike other types of eruptive star disks (e.g., EXors), FUors do not seem to show any crystalline silicate features. FU Orionis itself was found to be devoid of crystallines by \citet{2006ApJ...648..472Q}, and it would appear that the MATISSE observations corroborate the MIDI results.

\subsection{On a potential third companion}

\citet{2005A&A...437..627M} argued that FU~Ori might be a triple system, since modeling of their interferometric data ($H$ and $K$ bands) required an additional component from a point source (i.e., spot). The spot was estimated to be located at approximately 10~au (i.e., 25 mas) from the protostar and at a PA of about 130\degr.

\cite{2019ApJ...884...97L} reported a small deviation within the CO band in their GRAVITY/VLTI data, although the closure phase signal within errors is close to zero. Their geometric model fit to the data required an additional off-centered source, which they argue could be produced either by the presence of a nearby companion or by an inclined disk. 

\citet{2019ApJ...884...97L} and \citet{2020ApJ...889...59P} have shown that the CO gas in the disk is in Keplerian rotation, and therefore one could argue that the GRAVITY/VLTI observations probed the inner part of that gaseous component of the disk. Furthermore, \citet{2021A&A...646A.102L} argue that their closure phase measurements are null within error bars and, as such, more consistent with a centro-symmetric distribution that is inclined to our line of sight.

Similarly, we cannot confirm the presence of a third companion\footnote{The only confirmed companion, FU~Ori~South, is outside the field-of-view of the UT array.} from our data. At a first glance, none of the interferometric observables appear to show the sinusoidal signal expected from a binary companion. However, since this could be hidden inside the disk's signal, we attempted to test this. Essentially, faint companions at small separations would have produced a sinusoidal signal with a very low visibility amplitude and a very broad modulation cycle. We opted for a companion similar to the separation (25~mas) and flux ratio ($\leq3\%$) of \citet{2005A&A...437..627M} at multiple PAs, but find that if such a companion existed, its visibility amplitude ($\leq0.05$) would be undetectable with our current MATISSE $N$-band observations considering the large measurements in error ($\delta V\sim0.1$). We obtain similar results for companions with similar flux ratios but at larger separations (e.g., 100 mas).

Coincidentally, MATISSE can detect companions down to a flux ratio of 2\% in the $L$ band with the ATs, which offer a larger field-of-view, but the targets were wide binaries \citep[orbital separation $\sim 100$ mas; see][]{lopez2021matisse}. FU~Ori is marginally resolved in the $L$ band, while the visibilities are overall flat without showing any sinusoidal modulation (Fig.~\ref{fig:Lplots}). In a similar fashion to the approach presented above, we can exclude any companions in the separation range of $20-100$ mas with $L$-band flux ratios larger than 5\%.


\subsection{Investigating disk misalignments}\label{sec:misa}

Multiwavelength observations of protoplanetary disks have shown that under the influence of nearby companions (stellar or planetary), the inner regions of said disks can often become misaligned or warped. A striking example of disk misalignment is GW~Ori \citep{2020ApJ...895L..18B, 2020Sci...369.1233K}. Here, we investigate the possibility of disk misalignment for FU Orionis. 

Near-infrared scattered-light images \citep{2016SciA....2E0875L, 2018ApJ...864...20T,2020ApJ...888....7L} show a spiral-arc feature eastward of FU~Ori, while they suggest cavities in the north and northeastern directions. These large-scale ($\geq0.5$\arcsec) features could suggest an interaction with a companion, perhaps the known companion FU~Ori S, or even signify infalling material from the extended circumstellar environment onto the accretion disk.

Observations by ALMA \citep{2019ApJ...884...97L, 2020ApJ...889...59P} suggest a different orientation from any of the previous near-infrared, mid-infrared and radio interferometric observations (Table~\ref{tab:angles}). For example, \citet{2005A&A...437..627M} mention a best-fit model to their $H$- and $K$-band data, with an inclination of $\sim 55$\degr, as opposed to an earlier work \citep{1998ApJ...507L.149M} where a value of $\sim 30$\degr\ was measured in the $K$ band. Radio continuum observations at 33~GHz \citep[Karl G. Jansky Very Large Array  (JVLA); ][]{2017A&A...602A..19L} provide a highly uncertain disk PA, while the axis ratio suggests an inclination of $34^{+18}_{-8}$ deg, which is similar to the ALMA results.

Previously, \citet{2006ApJ...648..472Q} estimated from their MIDI/VLTI data an inclination similar to that of \citet{2005A&A...437..627M}; however, the disk's PA was quite different. Puzzling results can also be found in previous works, such as the reconstructed 10.7~\micron\ image of \citet{2009ApJ...700..491M} and the model image of \citet{2005A&A...437..627M}, which indicate that the disk's minor axis is oriented toward the northwest. \citet{2021A&A...646A.102L} reanalyzed archival $H$- and $K$-band interferometric observations and presented new $J$-band data from CHARA/MIRC-X. Their temperature gradient model suggests a disk with a PA$\sim 34$\degr\ and inclination of $\sim32$\degr\ (cf. Table~\ref{tab:angles}), which within errors agrees with \citet{2020ApJ...889...59P}.

We measured FU Ori disk's inclination and PA by fitting geometrical models to the $N$-band data. We attempted the same process for the GRA4MAT $L$-band data; however, since the disk is marginally resolved in this band, we focus more on the geometric properties derived from the $N$-band data. Within the uncertainties, the results agree very well with literature values (Table~\ref{tab:angles}). 

With the exception of the imaging results by \citet{2020ApJ...889...59P} in the submillimeter regime and by  \citet{2017A&A...602A..19L} in the radio regimes, all estimates for the disk's orientation were based on geometric model fits to interferometric data. In this work, we present radiative transfer models that used both the orientation derived by geometric fits to the MATISSE data and the orientation by ALMA \citep{2020ApJ...889...59P}. The latter appears to be more favorable, suggesting that the dusty disk is more pole-on in our line of sight. We cannot strongly conclude any disk misalignment for this system at present, especially for the hot accretion disk. We expect that even higher-angular-resolution instruments ($\theta\leq1$~mas) can provide more insight into that.


\begin{table}
        \caption{Accretion disk PA (minor axis) and inclination derived from interferometric data.}\label{tab:angles}
        \bgroup
    \def\arraystretch{1.5}%
        \begin{tabular}{lccc}
        \hline
        Reference & Band & PA (\degr) & $i$ (\degr) \\
        \hline\hline
        \citet{1998ApJ...507L.149M} & $K$       & -- & $\sim30$ \\
        Malbet et al. (2005) & $H,K$ & 47$^{+7}_{-11}$ & 55$^{+5}_{-7}$ \\
        \citet{2006ApJ...648..472Q} & $N$       & $19.1/3.4$ & $55.4\pm2.4$ \\
        \citet{2009ApJ...700..491M} & $N$ & $-25\pm35$ & -- \\
        Liu et al. (2017) & 33~GHz & $7.9\pm66$ & $34^{+18}_{-8}$ \\
        \citet{2020ApJ...889...59P} & 1.3~mm & $43.6\pm1.7$ & $37.7\pm0.8$ \\
        Labdon et al. (2021) & $J,H,K$ & $34\pm11$ & $32\pm4$ \\ \hline
        this work (UT; 2021) & $N$ & $15\pm25$ & $55\pm15$ \\
        this work (AT; 2021) & $L$ & $67\pm10$ & $48^{+7}_{-8}$ \\
        \hline
        \end{tabular}
}\end{table}

\section{Conclusions}\label{sec:conclusions}

We have presented new insights into the accretion disk of FU~Ori in the mid-infrared with the use of the interferometric instrument MATISSE/VLTI. Our results are complemented by radiative transfer simulations that attempt to constrain the properties of the disk.

In summary, we find that:
   \begin{itemize}
      \item The accretion disk is very compact in the $L$ band as it is marginally resolved at $<$2~mas in size, or 0.8~au at the adopted {\it Gaia} distance. Therefore, the hot accretion disk's radius ought to be smaller than 0.4~au. Similarly, MATISSE was able to detect just the innermost, and possibly warmer, part of the passive dusty disk, with an approximate size of 5~mas in the $N$ band, which translates to about 2 au.
      \item Geometric-model fits (2D Gaussian) to the MATISSE $N$-band correlated fluxes suggest a disk orientation with a minor-axis PA of $15\pm25$ degrees and an inclination of $55\pm15$ degrees, which are similar to literature values from MIDI/VLTI \citep{2006ApJ...648..472Q} but differ from more recent imaging studies by ALMA \citep{2020ApJ...889...59P}.
      \item Two radiative transfer models were explored for the two different disk orientations (MATISSE and ALMA). Both can provide relatively good fits to the SED and $L$-band data; however, a distinction can be seen when the models are compared with the $N$-band interferometric data (correlated fluxes and closure phases). It appears that an orientation that is more pole-on (i.e., ALMA) is more favorable. Since this discrepancy could allude to potential disk misalignment, we opt to reexplore this in future observations at even higher angular resolutions (e.g., GRAVITY/VLTI). 
      \item Our model fits suggest an average accretion rate of about $2\times10^{-5}$~$M_\sun$ yr$^{-1}$ (if we assume a stellar mass of 0.6~$M_\sun$), which is somewhat lower than literature values.
      \item There are no signatures of crystalline silicate features in either the total flux or the correlated $N$-band spectra, corroborating earlier results from MIDI/VLTI. The silicate feature itself is ``flat-topped,'' suggesting it emanates from large-sized grains (size$\geq2\,\rm\mu$m).
      \item Our model predicts a dust mass of $2.4\times10^{-4}\,M_{\sun}$ for the outer passive disk. A lower limit for the total disk mass (gas and dust) can be obtained by assuming a typical gas-to-dust ratio of 100, that is, $M^{\rm total}_{\rm disk}\sim 0.02$~$M_\sun$. However, we do not obtain any meaningful constraint on the mass of the inner gaseous disk other than it ought to be massive enough to be optically thick.
      \item The current MATISSE $N$-band observations cannot constrain the presence of an, as yet un-detected, tertiary companion for FU~Ori. In the $L$-band data, the absence of any sinusoidal modulation excludes any companions with a separation of $20-100$ mas and a flux ratio $\geq5\%$ in $L$.
   \end{itemize}
   
FU~Orionis, as the archetype of its class, is still an active laboratory for the study of post-eruptive accretion events, since it continues to fade in the optical and near-infrared and is not expected to return to its pre-eruption phase within the next 25 years.

We conclude that the MATISSE observations indicate that the hot, inner accretion disk is rather compact, with a diameter $\leq$1 au at 3.5$\rm\mu m$. Since earlier studies suggested a region of 1-2 au, this could indicate that the hot emitting region has been shrinking as the system fades.

Due to its low brightness and its compactness, FU~Orionis is not an ideal candidate for imaging in the mid-infrared, either with MATISSE or with aperture masking at larger spatial scales (e.g., VISIR/VLT). Imaging has also proven to be difficult with other near-infrared interferometers such as CHARA \citep{2021A&A...646A.102L}.

Modeling parameters could be constrained further by using observations in different array configurations than the ones presented here, should further advances in instrumentation allow observations of such faint targets. High-angular-resolution and high-spectral-resolution observations in the $K$ band (e.g., GRAVITY/VLTI) may be able to disentangle the gaseous and continuum emission at the center of the disk, and thus allow a combination of gas and dust modeling from the near- to the mid-infrared.

\begin{acknowledgements}
F.L. would like to thank the NOTCam team, and especially A.~A.~Djupvik, for their assistance in the planning and the qualitative analysis of the observations.

F.L. and P.A. are funded from the Hungarian NKFIH OTKA project no. K-132406. A.K., L.C., and M.S. are supported through ERC grant agreement No 716155 (SACCRED). J.V. is supported by NOVA, the Netherlands Research School for Astronomy. 

Our study is supported by the Lend\"ulet grant LP2012-31 of the Hungarian Academy of Sciences (MTA), and by the grant GINOP 2.3.2-15-2016-00033 of the National Research, Development and Innovation Office (NKFIH), Hungary, funded by the European Union.

Based on observations collected at the European Southern Observatory under ESO program 0104.C-0782(B), 0104.C-0016(D), 0106.C-0501(D) and 0106.C-0501(F), and on observations made with the Nordic Optical Telescope, owned in collaboration by the University of Turku and Aarhus University, and operated jointly by Aarhus University, the University of Turku and the University of Oslo, representing Denmark, Finland and Norway, the University of Iceland and Stockholm University at the Observatorio del Roque de los Muchachos, La Palma, Spain, of the Instituto de Astrofisica de Canarias.

MATISSE was designed, funded and built in close collaboration with ESO, by a consortium composed of institutes in France (J.-L.~Lagrange Laboratory – INSU-CNRS – C\^ote d’Azur Observatory – University of C\^ote d’Azur), Germany (MPIA, MPIfR and University of Kiel), the Netherlands (NOVA and University of Leiden), and Austria (University of Vienna). The Konkoly Observatory and Cologne University have also provided some support in the manufacture of the instrument. This research has made use of the services of the ESO Science Archive Facility.
\end{acknowledgements}

%
\bibliographystyle{aa} 
\bibliography{BIBFUOR}
%


\begin{appendix} 

\section{Contemporary SED}

The contemporary SED of FU Orionis, used as a comparison to the analytical and radiative transfer disk models, is listed in Table~\ref{tab:sed}. Here, we present our own photometric observations from SAAO, NOT and Konkoly Observatory (Sect.~\ref{sec:phot}), as well as archival mid- and far-infrared photometry obtained after 2010. The archival submillimeter continuum photometry was obtained after 2008. For the NEOWISE photometry in particular, we use data from the latest data release, and correct it for saturation following the instrument's documentation\footnote{\url{https://wise2.ipac.caltech.edu/docs/release/neowise/expsup/sec2_1civa.html}}.

\begin{table}[htbp]
    \centering
        \caption{FU Ori photometry, not corrected for extinction.}
    \label{tab:sed}
    \begin{tabular}{cccccc}\hline
    Band & $F_{\nu}$ (Jy) & $\sigma$ (Jy) & Source & Epoch & Ref.\\
    \hline\hline
        $U$    &  0.0305 & 0.0011 &  SAAO   & 2021 & t.w.\\
        $B$    &  0.1359 & 0.0014 &  SAAO   & \ldots &  t.w.\\
        $V$    &  0.4061 & 0.0037 &  SAAO   & \ldots &  t.w.\\
        $R_c$  &  0.6924 & 0.0082 &  SAAO   & \ldots &  t.w.\\
        $I_c$  &  1.2423 & 0.0125 &  SAAO   & \ldots &  t.w.\\
        $B$    &  0.1292 & 0.0014 &  RC80   & \ldots &  t.w.\\
        $V$    &  0.4338 & 0.0051 &  RC80   & \ldots &  t.w.\\
        $r'$   &  0.6841 & 0.0083 &  RC80   & \ldots &  t.w.\\
        $i'$   &  1.1341 & 0.0172 &  RC80   & \ldots &  t.w. \\
    $G_{\rm BP}$  & 0.3322      & 0.0011  &  {\it Gaia} DR2 & 2015 & (1) \\
    $G$  & 0.5811       & 0.0008  &  {\it Gaia} DR2 & 2015 & (1) \\
    $G_{\rm RP}$  & 1.1387 & 0.0047  & {\it Gaia} DR2 & 2015 & (1) \\
    $J$   & 2.9491      & 0.0751 & NOT & 2021 &  t.w. \\
    $H$   & 3.5083      & 0.0988 & NOT & \ldots &  t.w. \\
    $Ks$  & 4.0031      & 0.0803 & NOT & \ldots &  t.w. \\ 
    $W1$        & 3.7126 &      0.2542 & NEOWISE & 2020 & (2) \\
    $W2$        & 3.3039 &      0.2431 & NEOWISE & \ldots & (2) \\\hline
    $\rm \lambda\,(\mu m)$ & $F_{\nu}$ (Jy) & $\sigma$ (Jy) & Source & Epoch & Ref.\\ \hline\hline
    5.6   & 4.24 & 2.06 & SOFIA  & 2016 & (3) \\
    7.7   & 3.64 & 1.91 & SOFIA  & \ldots & (3) \\
    11.1  & 4.21 & 0.00 & SOFIA  & \ldots & (3) \\
    25.3  & 7.68 & 2.79 & SOFIA  & \ldots & (3) \\
    31.5  & 8.74 & 2.96 & SOFIA  & \ldots & (3) \\
    34.8  & 8.54 & 2.94 & SOFIA  & \ldots & (3) \\
    70  & 6.5 & 0.7     & {\it Herschel} & 2012 &  (4) \\
        100     & 5.9 & 0.6     & {\it Herschel} & \ldots &  (4) \\
        160     & 4.9 & 0.5     & {\it Herschel} & \ldots &  (4) \\
        250     & 2.8 & 0.5     & {\it Herschel} & 2011 &  (4) \\
        350     & 2.5 & 0.6     & {\it Herschel} & \ldots &  (4) \\
    500  & 0.9 &        0.3  & {\it Herschel} & \ldots &  (4) \\
    853  & 0.0501 &     $3e-04$ & ALMA & 2012 & (5) \\
    1100  & 0.0350 &    0.0070 &        SMA & 2008 & (6) \\
    1300  & 0.0143 &    0.0015 &        ALMA & 2017 & (7) \\
    \hline
    \end{tabular}
    \tablebib{``t.w.'' = this work; (1) \citet{gaiaDR2paper}; (2) corrected for saturation; (3) \citet{green2016}; (4) \citet{green2013}; (5) \citet{2015ApJ...812..134H}; (6) \citet{2017A&A...602A..19L} ; (7) \citet{2020ApJ...889...59P}}
\end{table}

\section{Observations log}\label{sec:log}

The observation log for all five epochs is presented in Table~\ref{tab:log}, where DIT is the integration time, and $\tau_0$ is the atmospheric coherence time. A short description of the data quality for all five epochs is given below.

\paragraph{Epoch 1:} The detection limits for the AT configurations in low-spectral-resolution mode are 1, 5, and 10~Jy, respectively, for the $L$, $M$, and $N$ bands. We found that FU Ori was fainter than expected, that is, $\leq 5$~Jy in all three bands, so the first attempt to observe it with an AT configuration was not successful. Although the star was slightly brighter than the $L$-band limit, the data were of low signal-to-noise. The $N$-band data are of very poor quality. We therefore reject these observations.

\paragraph{Epoch 2:} Data were obtained in adverse atmospheric conditions with seeing as high as 1.5\arcsec\ and atmospheric coherence time $\leq4$~ms. This affected mostly the $L$ and $M$ observations as opposed to the $N$ band; however, we also found problems with the absolute photometric performance of the UTs in $N$ (total flux spectra). This data set was used only as qualitative comparison for epoch 4 $N$-band data.

\paragraph{Epoch 3:} No matching calibrator was observed in this epoch. Therefore, that data set is not included in this work.

\paragraph{Epochs 4 and 5:} Technical issues in epoch 4 hindered the observations. The designated calibrator (HD37160) was found to be brighter than expected and saturated; therefore, an additional calibrator (HD47886) was observed immediately after. However, another technical error occurred, and the connection to UT2 telescope was lost; therefore, only three of the six baselines could be calibrated for that run (HD47886). We therefore opted to recalibrate epoch 4 data with the first calibrator of that night's GTO program (HD28413), which were obtained at a similar airmass as FU Ori. In general, we find that the chopped interferometric data are of better quality in $L$ and $M$, since temporal variations in atmospheric transmission in the mid-infrared could be corrected with chopping. We also find that the total spectra, the correlated fluxes, and the visibilities from this third calibrator agree within 10\% with the results from HD47886, and moreover the total spectra are nearly identical to those obtained with a different mode in epoch 5. Therefore, here we show the data sets calibrated with HD28413. We also note that all $N$-band data obtained here in high-spectral-resolution mode suffer from poor S/N beyond 11\micron. This may be the result of incorrect flat-fielding, manifesting as correlated noise features, because this spectral mode utilizes a larger portion of the detector that has not been characterized well yet\footnote{At the time this work was submitted to the journal, delays due to the COVID-19 pandemic had hindered the optimization of the instrument.}.

\paragraph{Epoch 5*:} In addition to the low-spectral-resolution data in $LM$ for epoch 5, we obtained data in medium-spectral resolution in $LM$ and tested the performance of the high-spectral-resolution mode in $N$. It proved unsuccessful since the brightness limit for that mode is found to be 26~Jy for the correlated fluxes. The $L$-band medium-spectral resolution data are beyond the scope of this work and will be analyzed in a future publication.

\begin{table*}
\caption{Observing log.}\label{tab:log}
\begin{tabular}{ccccccccc}
\hline
Epoch & Date & Band & $R$ & DIT & Configuration & Seeing & $\tau_0$ & Calibrator\\
& & & ($\lambda/\Delta\lambda$) & (ms)  & array & (\arcsec) & (ms) & \\ 
 \hline\hline
1 & 2019-12-22T04:51:39 & LM & low & 111 & K0-G2-D0-J3  & 0.46 & 7.83 & HD49161 \\
 & " & N  & low & 20  & K0-G2-D0-J3  & 0.53 & 8.56 & HD49161  \\ \hline
2 & 2020-03-15T00:23:24 & LM & low & 111 & U1-U2-U3-U4  & 1.04 & 3.14 & HD48433 \\
 & " & N  & low & 20  & U1-U2-U3-U4  & 1.10 & 2.90 & HD48433 \\ \hline
3 & 2021-01-07T04:40:21 & LM & low & 111 & U1-U2-U3-U4 & 0.71 & 5.23 & --\\
 & " & N & high & 75 & U1-U2-U3-U4 & 0.71 & 5.23 & -- \\ \hline
4 & 2021-01-08T03:33:01 & LM & low & 111 & U1-U2-U3-U4 & 0.63 & 7.87 & HD28413, HD47886 \\
 & " & N & high & 75 & U1-U2-U3-U4 & 0.55 & 10.12 & HD28413, HD47886\\ \hline
5* & 2021-01-16T02:36:10 & LM & med & 3000 & A0-G1-J2-J3 & 0.84 & 7.31 & HD31767 \\
 & 2021-01-16T02:41:21 & N & high & 75 & A0-G1-J2-J3 & 0.81 & 7.32 & HD31767 \\ \hline
5 & 2021-01-16T02:47:06 & LM & low & 111 & A0-G1-J2-J3 & 0.70 & 7.30 & HD31767 \\
 & " & LM & low & 111 & A0-G1-J2-J3 & 0.66 & 8.42 & HD31767 \\
\hline
\end{tabular}
\end{table*}

\FloatBarrier
\section{Extended model comparison}
\begin{figure*}
        \centering
                \includegraphics[scale=0.65]{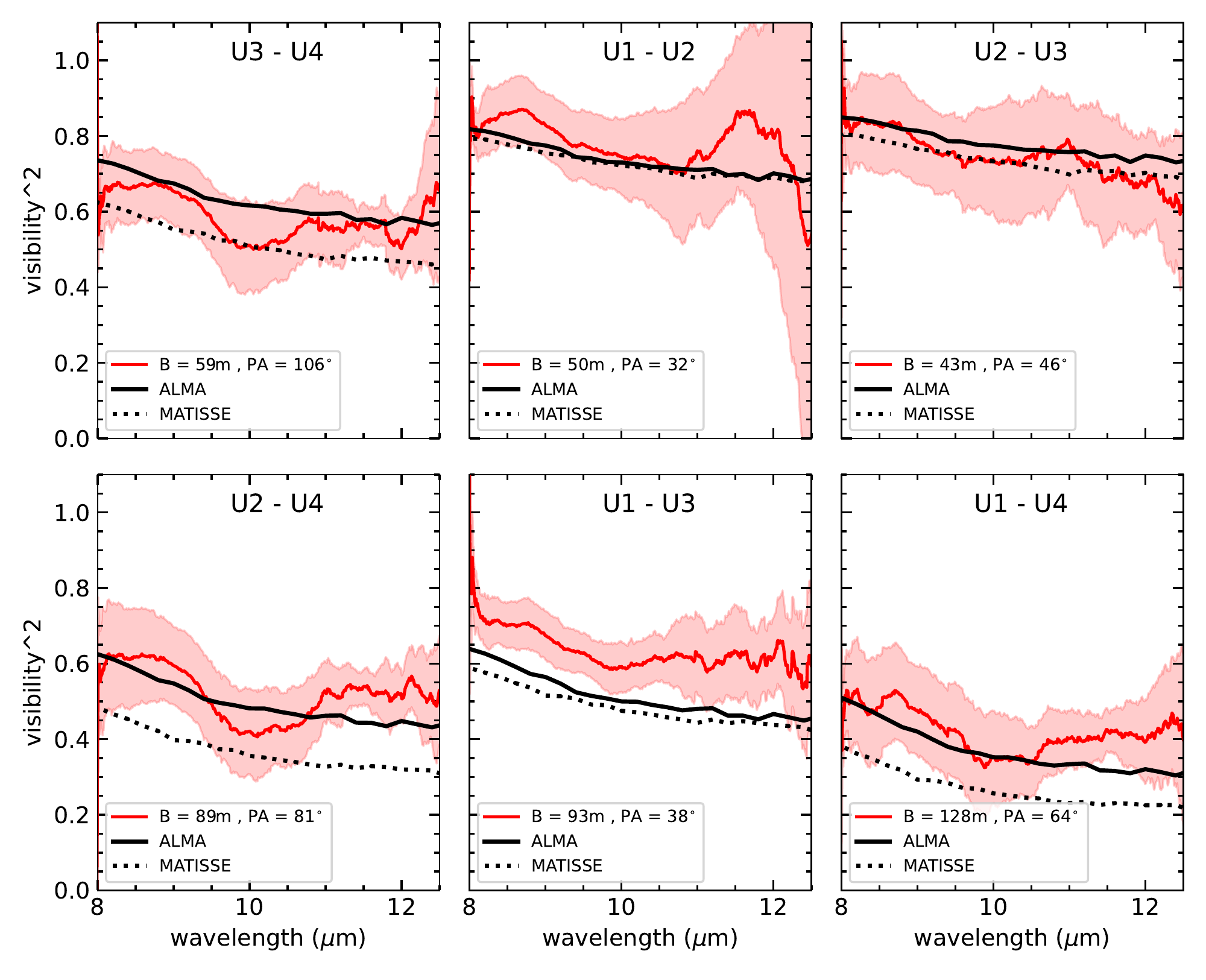}
        \caption{Synthetic visibilities (squared) for the two {\sc radmc-3d} models at two disk orientations (ALMA, solid black; MATISSE, dashed) against the MATISSE $N$-band data. }
        \label{fig:extra_vis2}
\end{figure*}

\begin{figure*}
        \centering
        \includegraphics[scale=0.65]{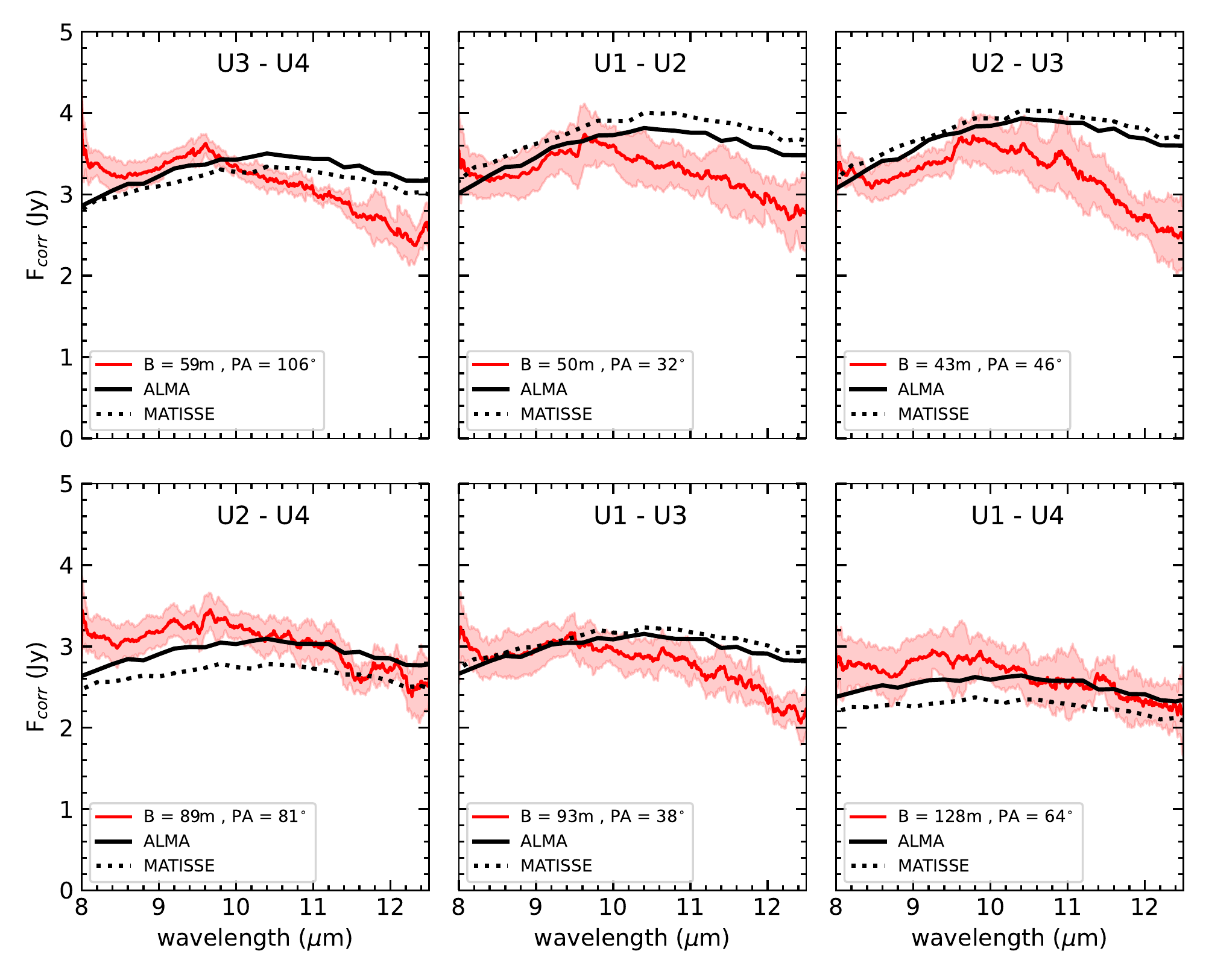}
        \caption{As in Fig.~\ref{fig:extra_vis2}, but the correlated fluxes. }
        \label{fig:extra_fcorr}
\end{figure*}

Further comparisons of the two {\sc radmc-3d} models against the MATISSE $N$-band data are provided in Figs.~\ref{fig:extra_vis2} and \ref{fig:extra_fcorr}, which show the synthetic visibilities and correlated fluxes with respect to wavelength, respectively. Both models can fit the visibilities within observational uncertainties, but the ALMA-orientation model provides better fits for the correlated fluxes. 

\end{appendix}
\end{document}